\input jytex.tex   
\typesize=10pt
\magnification=1200
\baselineskip17truept
\footnotenumstyle{arabic}
\hsize=6truein\vsize=8.5truein
\sectionnumstyle{blank}
\chapternumstyle{blank}
\chapternum=1
\sectionnum=1
\pagenum=0

\def\begintitle{\pagenumstyle{blank}\parindent=0pt\begin{narrow}[0.4in]}
\def\endtitle{\end{narrow}\newpage\pagenumstyle{arabic}}


\def\beginexercise{\vskip 20truept\parindent=0pt\begin{narrow}[10
truept]}
\def\endexercise{\vskip 10truept\end{narrow}}


\def\eql#1{\eqno\eqnlabel{#1}}
\def\ref{\reference}
\def\peq{\puteqn}
\def\pref{\putref}

\def\mgn{\marginnote}
\def\bex{\begin{exercise}}
\def\eex{\end{exercise}}


\font\open=msbm10 
\font\goth=eufm10  
\font\ssb=cmss10
\def\mbox#1{{\leavevmode\hbox{#1}}}

\def\hspace#1{{\phantom{\mbox#1}}}
\def\oR{\mbox{\open\char82}}

\def\oZ{\mbox{\open\char90}}

\def\gK{\mbox{{\goth\char75}}}

\def\ssx{\mbox{{\ssb\char120}}}
\def\sssb{\mbox{{\ssb\char98}}}

\def\al{\alpha}
\def\be{\beta}
\def\ga{\gamma}
\def\de{\delta}
\def\Ga{\Gamma}

\def\ep{\epsilon}

\def\ka{\kappa}
\def\la{\lambda}

\def\om{\omega}

\def\si{\sigma}

\def\th{\theta}
\def\Th{\Theta}
\def\ze{\zeta}

\def\De{\Delta}

\def\det{{\rm det\,}}

\def\Real{{\rm Re\,}}

\def\zf{$\zeta$--function}
\def\zfs{$\zeta$--functions}


\def\frac#1/#2{\leavevmode\kern.1em
\raise.5ex\hbox{\the\scriptfont0 #1}\kern-.1em/\kern-.15em
\lower.25ex\hbox{\the\scriptfont0 #2}}
\def\sfrac#1/#2{\leavevmode\kern.1em
\raise.5ex\hbox{\the\scriptscriptfont0 #1}\kern-.1em/\kern-.15em
\lower.25ex\hbox{\the\scriptscriptfont0 #2}}

\def\gtorder{\mathrel{\raise.3ex\hbox{$>$}\mkern-14mu
             \lower0.6ex\hbox{$\sim$}}}
\def\ltorder{\mathrel{\raise.3ex\hbox{$<$}\mkern-14mu
             \lower0.6ex\hbox{$\sim$}}}

\def\semidirprod{\rlap{\ss C}\raise1pt\hbox{$\mkern.75mu\times$}}
\def\for{\lower6pt\hbox{$\Big|$}}
\def\fish{\kern-.25em{\phantom{abcde}\over \phantom{abcde}}\kern-.25em}


\def\boxit#1{\vbox{\hrule\hbox{\vrule\kern3pt
        \vbox{\kern3pt#1\kern3pt}\kern3pt\vrule}\hrule}}
\def\dalemb#1#2{{\vbox{\hrule height .#2pt
        \hbox{\vrule width.#2pt height#1pt \kern#1pt
                \vrule width.#2pt}
        \hrule height.#2pt}}}

\def\ol{\overline}

\def\frac#1#2{{{#1}\over{#2}}}

\def\comb#1#2{{\left(#1\atop#2\right)}}

\def\cosec{{\rm cosec\,}}

\def\eg{{\it e.g. }}
\def\ie{{\it i.e. }}
\def\cf{{\it cf }}
\def\pa{\partial}


  %

\def\sumdasht#1#2{{\mathop{{\sum}'}_{#1}^{#2}}}

\def\3j#1#2#3#4#5#6{\left\lgroup\matrix{#1&#2&#3\cr#4&#5&#6\cr}
\right\rgroup}

\def\caS{{\cal S}}

\def\man{{\cal M}}

\def\caD{{\cal D}}
\def\caF{{\cal F}}

\def\m?{\mgn{?}}

\def\pa{\partial}

\def\beq{\begin{eqnarray}}
\def\eeq{\end{eqnarray}}


\def\cqg#1#2#3{{\it Class. Quant. Grav.} {\bf {#1}} ({#2}) #3}

\def\jmp#1#2#3{{\it J. Math. Phys.} {\bf {#1}} ({#2}) #3}
\def\jpa#1#2#3{{\it J. Phys.} {\bf A{#1}} ({#2}) #3}

\def\np#1#2#3{{\it Nucl. Phys.} {\bf B{#1}} ({#2}) #3}
\def\pl#1#2#3{{\it Phys. Lett.} {\bf {#1}} ({#2}) #3}

\def\prp#1#2#3{{\it Phys. Rep.} {\bf {#1}} ({#2}) #3}
\def\pr#1#2#3{{\it Phys. Rev.} {\bf {#1}} ({#2}) #3}

\def\prD#1#2#3{{\it Phys. Rev.} {\bf D{#1}} ({#2}) #3}

\def\prs#1#2#3{{\it Proc. Roy. Soc.} {\bf A{#1}} ({#2}) #3}

\def\amsh#1#2#3{{\it Abh. Math. Sem. Ham.} {\bf {#1}} ({#2}) #3}
\def\am#1#2#3{{\it Acta Mathematica} {\bf {#1}} ({#2}) #3}
\def\aim#1#2#3{{\it Adv. in Math.} {\bf {#1}} ({#2}) #3}

\def\aom#1#2#3{{\it Ann. of Math.} {\bf {#1}} ({#2}) #3}

\def\dmj#1#2#3{{\it Duke Math. J.} {\bf {#1}} ({#2}) #3}
\def\invm#1#2#3{{\it Invent. Math.} {\bf {#1}} ({#2}) #3}

\def\jram#1#2#3{{\it J. f. reine u. Angew. Math.} {\bf {#1}} ({#2}) #3}
\def\jims#1#2#3{{\it J. Indian. Math. Soc.} {\bf {#1}} ({#2}) #3}
\def\jlms#1#2#3{{\it J. Lond. Math. Soc.} {\bf {#1}} ({#2}) #3}

\def\ma#1#2#3{{\it Math. Ann.} {\bf {#1}} ({#2}) #3}

\def\mz#1#2#3{{\it Math. Zeit.} {\bf {#1}} ({#2}) #3}

\def\plms#1#2#3{{\it Proc. Lond. Math. Soc.} {\bf {#1}} ({#2}) #3}
\def\pgma#1#2#3{{\it Proc. Glasgow Math. Ass.} {\bf {#1}} ({#2}) #3}
\def\qjm#1#2#3{{\it Quart. J. Math.} {\bf {#1}} ({#2}) #3}

\def\rmjm#1#2#3{{\it Rocky Mountain J. Math.} {\bf {#1}} ({#2}) #3}

\def\tams#1#2#3{{\it Trans.Am.Math.Soc.} {\bf {#1}} ({#2}) #3}

\vglue 1truein
\vskip15truept
\centertext{\Bigfonts \bf Modular properties of Eisenstein}\vskip10truept
\centertext {\Bigfonts \bf series and statistical mechanics}
\vskip10truept \centertext{\Bigfonts \bf }

 \vskip 20truept
\centertext{J.S.Dowker\footnote{dowker@man.ac.uk} } \vskip 7truept
\centertext{\it Theoretical Physics Group, } \centertext{\it School of
Physics and Astronomy}\centertext{\it The University of Manchester,}
\centertext{ \it Manchester, England} \vskip 10truept


 \vskip15truept
The temperature inversion properties of the internal energy, $E$, on odd
spheres, and its derivatives, together with their expression in elliptic
terms, as expounded in previous papers, are extended to the {\it
integrals} of $E$, thence making contact with the theory of modular forms
with rational period functions.

I point out that the period functions of (holomorphic) Eisenstein series
computed by Zagier were already available since the time of Ramanujan and
I give a rederivation by contour integration. Removing both the Planck
and Casimir terms gives a fully subtracted form of the series which
allows a more elegant and compact treatment. I expound the relation to
Eichler cohomology cocycles and also rewrite the theory in a
distributional, Green function way.

Some historical and technical developments of the Selberg--Chowla formula
are presented, and it is suggested that this be renamed the
Epstein--Kober formula. On another point of historical justice, the work
of Koshliakov on Dirichlet series is reprised. A representation of a
`massive' generalised Dirchlet series due to Berndt is also reproved,
applied to the Epstein series and to a derivation of the standard
statistical mode sum, interpreted as a Kronecker limit formula.

\newpage
\section{\bf 1. Introduction.}

In two previous works, [\pref{DandK2}], [\pref{DandK1}], we have
investigated the thermal quantities for scalar and spinor fields on the
space--time T$\times$S$^d$, where $d$ is odd. The conformal scalar
internal energy, $E$, for example, is a linear combination of `partial'
energies, $\ep_t,\, t=1,2,\ldots$, (see below). These quantities possess
a known behaviour under modular transformations and are central to
elliptic function theory. In particular, a temperature inversion symmetry
can be exhibited.

From basic elliptic properties, it was shown that $E$ could be expressed
as a polynomial in $\ep_2$ and $\ep_3$, while, for the specific heat, and
all higher derivatives, one has to include $\ep_1$, preferably via a
modular covariant derivative. The free energy involves an integration and
things are not so simple as obstructions arise to simple modular
behaviour. It is this aspect that is explored in the present work. I use
this thermal angle just to motivate the introduction of various
quantities, the analysis of which allows us to make contact with various
topics in the theory of modular forms. Some of the procedures can be
given a physical terminology, which might be suggestive.

\section{\bf 2. Mellin transforms and the period polynomial.}

It is advantageous to begin from the basic Mellin transform, used \eg by
Malurkar and Hardy, for the Eisenstein series (or partial internal
energy, [\pref{DandK1}] [\pref{DandK2}]),
  $$\eqalign{
  \ep_t({\sssb})&=-{B_{2t}\over 4t}+
  \sum_{n=1}^\infty {n^{2t-1}q^{2n}\over1-q^{2n}}\cr
   &={1\over2}\,\ze_R(1-2t)+{1\over2\pi i}\int_{c-i\infty}^{c+i\infty}ds\,\Ga(s)\,
  \ze_R(s)\,\ze_R(s-2t+1)\,(2\pi{\sssb})^{-s}\,,}
  \eql{mell1}$$
where $q=e^{i\pi\tau}$ and $c$ lies above $2t$. I am using
${\sssb}=-i\tau= \be/2\pi=1/\xi$ as yet another convenient variable.
Although, physically, $\sssb$ is real, if we wish to consider general
modular transformations then we must allow it to become complex.

The Mellin transform has been used systematically in finite temperature
field theory and elsewhere in discussions of asymptotic limits. The
survey by Elizalde {\it et al}, [\pref{Elizalde2}], is useful and I
further mention, as being somewhat relevant here, Cardy [\pref{Cardy}],
Kutasov and Larsen, [\pref{KandL}]. Terras, [\pref{Terras}] pp.55,229,
can be consulted for aspects of the mathematical side. \mgn{OTHERS} In
number theory see Hardy and Littlewood, [\pref{HandL}], and Landau,
[\pref{Landau,Landau2}].

Equation (\peq{mell1}) can be looked upon as a continuous expansion in
${\sssb}$. Various limits in ${\sssb}$ are uncovered by displacing the
integration contour. The pole in the integrand at $s=2t$ corresponds to
the Planck term, while the remaining one at $s=0$ corresponds to minus
the Casimir value, which is the first term in (\peq{mell1}). The high
temperature limit (${\sssb}\to0$) follows by pushing the contour to the
left, beyond the Casimir pole, the contribution from which cancels the
first term in (\peq{mell1}) leaving just the effect of the Planck pole
and a correction that tends to zero exponentially. Conversely, on moving
the contour all the way to the right only the zero temperature Casimir
term remains.

This cosmetic asymmetry can be avoided (if desired) by also extracting
the Planck term, defining,
  $$
  \ep_t^{\rm sub}({\sssb})=\ep_t(\sssb)-{1\over2}\,\ze_R(1-2t)
  \big(1+({i\sssb})^{-2t}\big)\,,
  \eql{rege}$$
and writing
  $$\eqalign{
\ep_t^{\rm sub}({\sssb})={1\over2\pi i}
\int_{c-i\infty}^{c+i\infty}ds\,\Ga(s)\,
  \ze_R(s)\,\ze_R(s-2t+1)\,(2\pi{\sssb})^{-s}\,,}
  \eql{mell2}$$
where now $c$, at the moment, lies anywhere {\it between} $0$ and $2t$.

The quantity in the integrand is the Hecke $L$--series of the normalised
Eisenstein modular form,
  $$
  L(G_t,s)=\ze_R(s)\,\ze_R(s-2t+1)\,,
  \eql{eisl1}$$
satisfying the (typical) functional relation
  $$
  (2\pi)^{-s}\Ga(s)\,L(G_t,s)=(-1)^t(2\pi)^{s-2t}\Ga(2t-s)\,L(G_t,2t-s)\,.
  \eql{gfr}$$

I now consider a multiple integration with respect to ${\sssb}$ of the
internal energy. Actually for the free energy only one integration is
needed but the general case is mathematically important. So integrate
$\ep_t^{\rm sub}$, $h$ times to give, using (\peq{mell2}),
  $$\eqalign{
{1\over\Ga(h)}&\int_\ssx^\infty d{\sssb}\,({\sssb}-\ssx)^{h-1}
\,\ep_t^{\rm sub}({\sssb})\cr
 & ={1\over2\pi i}
\int_{c-i\infty}^{c+i\infty}ds\,{\Ga(s-h)\over(2\pi)^s}\,
  \ze_R(s)\,\ze_R(s-2t+1)\,\ssx^{-s+h}\,,}
  \eql{mint}$$
with $h<c<2t$.

I have imposed the boundary condition that each successive integral
vanishes at $\ssx=\infty$, \ie at zero temperature.

More poles have been introduced into the integrand from the $\Ga$
function and the contour has to be moved to the right to avoid them but
we should stop when they pass beyond the contour at $ \Real
s=c=2t-\de,\,0<\de<1$. Hence the maximum $h$ can be is $2t-1$.

I now set $h$ equal to this maximum, $2t-1$, simply for the reason that
it gives the ultimately recognisable, and important, quantity,
  $$\eqalign{
{\overline\phi}_{2t}(\ssx)&\equiv{2(2\pi)^{2t}\over\Ga(2t-1)}
\int_\ssx^\infty d{\sssb}\,({\sssb}-\ssx)^{2t-2}\, \ep_t^{\rm
sub}({\sssb})\cr
 & ={2\over2\pi i}
\int_{c-i\infty}^{c+i\infty}ds\,{\Ga(s-2t+1)\over(2\pi)^{s-2t}}\,
  \ze_R(s)\,\ze_R(s-2t+1)\, \ssx^{-s+2t-1}\,,}
  \eql{mint2}$$
with $2t-1<c<2t$. The first line is recognised as a Weyl fractional
integral.

The significance of this quantity for statistical mechanics on spheres is
admittedly obscure because the internal energy is a sum of the $\ep_t$
for different $t$, and the integral depends on $t$. Only for the
3--sphere, where $t=2$ and $\overline E=\ep_2$, is there a direct
relation but even then $\overline \phi_4$ is a third integral of $E$
having no particular thermal meaning. Nevertheless, its further analysis
is not without interest as it connects with some salient mathematical
concepts.

According to my general programme, interest lies in the behaviour under
modular transformations. My treatment is equivalent to that of Malurkar,
[\pref{Malurkar}], \cf also Guinand [\pref{Guinand}]. In contrast to the
derivatives of $E$, there are obstructions to the modular invariance of
the integrals and the {\it period polynomials} or, rather, period {\it
functions}, provide a measure of these obstructions, as we will see.

Firstly under translations, ${\sssb}\to{\sssb}-i$, and, because of the
Planck subtraction,
  $$ \ep_{t}^{\rm
sub}({\sssb}-i)-\ep_{t}^{\rm
sub}({\sssb})=(-)^{t+1}{1\over2}\,\ze_R(1-2t)
\bigg({1\over({\sssb}-i)^{2t}}- {1\over{\sssb}^{2t}}\bigg)\,,
  \eql{trans1}$$
which implies, through (\peq{mint2}), that
  $$
  {\overline\phi}_{2t}(\ssx-i)-{\overline\phi}_{2t}(\ssx)
  =2i{\ze_R(2t)\over \ssx(\ssx-i)}\,.
  \eql{trans}$$

Under inversion, $\ssx\to1/\ssx$, the integrand involves
  $$
  \ep_t^{\rm sub}(1/{\sssb})=(-1)^t {\sssb}^{2t}\,\ep^{\rm sub}_t({\sssb})
  $$
since the regularising subtraction, (\peq{rege}), maintains the inversion
property,
  $$
  \ep_t(1/{\sssb})=(-1)^t {\sssb}^{2t}\,\ep_t({\sssb})\,,
  $$
enjoyed by the full Eisenstein series in (\peq{mell1}),

Therefore,

$$\eqalign{
  {\overline\phi}_{2t}(1/\ssx)=&{2(2\pi)^{2t}\over\Ga(2t-1)}\int_{1/\ssx}^{\infty}
  d\sssb\,({\sssb}-1/\ssx)^{2t-2}\,\ep_t^{\rm sub}({\sssb})\cr
  =&-{2(2\pi)^{2t}\over\Ga(2t-1)}\int_\ssx^0
  {d{\sssb}'\over{\sssb}'^2}\,\big(1/{\sssb}'-1/\ssx\big)^{2t-2}\,
\ep^{\rm sub}_t(1/{\sssb}')\cr
  =&(-1)^t\,\ssx^{2-2t}{2(2\pi)^{2t}\over\Ga(2t-1)}\int_0^\ssx
  d{\sssb}\,(\ssx-{\sssb})^{2t-2}\,\ep^{\rm sub}_t({\sssb})\,.
  }
  \eql{mint3}
  $$

The amount by which ${\overline\phi}_{2t}$ violates the pure modular
inversion property is the quantity,
  $$
  {\overline\phi}_{2t}(\ssx)-(i\ssx)^{2t-2}\,{\overline\phi}_{2t}(1/\ssx)
\equiv \overline P_t(\ssx)\,,
  \eql{pee1}$$
or, from (\peq{mint2}) and (\peq{mint3})
  $$\eqalign{
  \overline P_t(\ssx)=&{2(2\pi)^{2t}\over\Ga(2t-1)}
  \int_0^{\infty}d{\sssb}\,\,(\ssx-{\sssb})^{2t-2}\,
\ep^{\rm sub}_t({\sssb})}\,.
  \eql{ppol}$$

The calculation of $\overline P_t$ from the Mellin contour form goes as
follows. From (\peq{mint2}),
  $$\eqalign{
&{\overline\phi}_{2t}(1/\ssx)\cr&={2\over2\pi i}
\int_{c-i\infty}^{c+i\infty}ds\,{\Ga(s-2t+1)\over(2\pi)^{s-2t}}\,
  \ze_R(s)\,\ze_R(s-2t+1)\,\ssx^{s-2t+1}\cr
&={2\over \ssx^{2t-2}}{1\over2\pi i}
\int_{2t-c-i\infty}^{2t-c+i\infty}ds\,{\Ga(1-s)\over(2\pi)^{-s}}\,
  \ze_R(2t-s)\,\ze_R(1-s)\,\ssx^{2t-s-1}\,,}
  \eql{mint4}$$
by setting $s\to2t-s$.

The $\ze$--functional equation (one can also, more easily, use
(\peq{gfr})) leads to,
  $$\eqalign{
  &\ze_R(s)\,\ze_R(s-2t+1)\cr&=
  {2\Ga(1-s)\over(2\pi)^{1-s}}\sin(\pi s/2)\,\ze_R(1-s){2\Ga(2t-s)
  \over(2\pi)^{2t-s}}
  \sin(\pi(s-2t+1)/2)\ze_R(2t-s)\cr
  &={4\Ga(1-s)\Ga(2t-s)\over(2\pi)^{2t+1-2s}}\sin(\pi s/2)\,
  \sin(\pi(s-2t+1)/2)\ze_R(1-s)\ze_R(2t-s)\cr
   &=(-1)^t{2\Ga(1-s)\Ga(2t-s)\over(2\pi)^{2t+1-2s}}\sin(\pi s)
   \,\ze_R(1-s)\ze_R(2t-s)\cr
    &={(-1)^{t-1}\Ga(1-s)\over(2\pi)^{2t-2s}\Ga(s-2t+1)}\,\ze_R(1-s)\ze_R(2t-s)\,,
  }
  $$
and so (\peq{mint4}) becomes,
  $$\eqalign{
&{\overline\phi}_{2t}(1/\ssx)\cr
 &={2(-1)^{t-1}\over \ssx^{2t-2}}{1\over2\pi i}
\int_{2t-c-i\infty}^{2t-c+i\infty}ds\,{\Ga(s-2t+1)\over(2\pi)^{s-2t}}\,
  \ze_R(s)\,\ze_R(s-2t+1)\,\ssx^{2t-s-1}\,.}
  \eql{mint5}$$
I then find the alternative expression for the obstruction, $\overline
P_t(\ssx)$, defined in (\peq{pee1}) or (\peq{ppol}),
  $$
  \overline P_t(\ssx)={1\over\pi i}\bigg(
\int_{c-i\infty}^{c+i\infty}-
\int_{2t-c-i\infty}^{2t-c+i\infty}\bigg)ds\,{\Ga(s-2t+1)\over(2\pi)^{s-2t}}\,
  \ze_R(s)\,\ze_R(s-2t+1)\,\ssx^{2t-s-1}\,.
  $$
Translating the contours so that they mutually cancel leaves the
contributions of the intervening poles \ie those between 0 and $2t$. Thus
  $$\eqalign{
  \overline P_t(\ssx)&={1\over\pi i}\oint_C ds\,{\Ga(s-2t+1)
  \over(2\pi)^{s-2t}}\,
  \ze_R(s)\,\ze_R(s-2t+1)\,\ssx^{2t-s-1}\cr
  &=(-1)^{t-1}{1\over i}\oint_C ds\,
  \ze_R(s)\,\ze_R(2t-s)\,\cosec (\pi s/2)\ssx^{2t-s-1}\cr
  &=2\pi\ze_R(2t-1)\big((i \ssx)^{2t-2}-1\big)-4i
  \sum_{j=1}^{t-1}\ze_R(2j)\,\ze_R(2t-2j)(i \ssx)^{2t-2j-1}\cr
  &=2\pi\ze_R(2t-1)\big((i \ssx)^{2t-2}-1\big)-i(-1)^t(2\pi)^t
  \sum_{j=1}^{t-1}{B_{2j}\,B_{2t-2j}\over (2j)!(2t-2j)!}\,(i \ssx)^{2t-2j-1}
  \,,
  }
  \eql{ppoly2}$$
which is a polynomial of degree $(2t-2)$ and is the final, main result of
this section. The odd powers come from the poles of the cosec while the
two even powers arise from the poles of the Riemann \zfs. $\overline
P_t(\ssx)$ is real if $\ssx$ is.

Setting $\ssx$ to zero in (\peq{ppoly2}) and (\peq{ppol}) produces the
specific formulae,
  $$\eqalign{
  \ze_R(2t-1)&=-{2(2\pi)^{2t-1}\over\Ga(2t-1)}\int_0^\infty d\sssb\,
  \sssb^{2t-2}\ep_t^{\rm sub}(\sssb)\cr
  &=-(-1)^t{2(2\pi)^{2t-1}\over\Ga(2t-1)}\int_0^\infty d\sssb\,
  \ep_t^{\rm sub}(\sssb)\,,
  }
  \eql{even}
  $$
using inversion. These can be checked numerically. More generally,
expanding (\peq{ppol}) gives the other nonzero moments of $\ep_t^{\rm
sub}(\sssb)$,
  $$
  \int_0^\infty d\sssb\,\sssb^{2j-1}\ep_t^{\rm
  sub}(\sssb)={(-1)^j\over8j(t-j)}\,B_{2j}\,B_{2t-2j}\,,\quad 1<j<t-1\,.
  \eql{moments}
  $$
The periods given by Kohnen and Zagier, [\pref{KandZ}], are identical to
(\peq{even}) and (\peq{moments}) and I now have made contact with this
mathematical notion.

The {\it periods} of a {\it cusp} modular form, $\caF$, of weight $2t$
are the numbers (\eg\ Lang [\pref{Lang}]),\footnote{ I prefer the
normalisation of Kohnen and Zagier, [\pref{KandZ}].}
  $$
  r_n(\caF)=\int_0^{\infty}d\sssb\,\caF(i\sssb)
  \,\sssb^n\,,\quad 0\le n\le 2t-2\,.
  \eql{periods}$$
The end points are cusps where $\caF$ vanishes (exponentially) to ensure
convergence. The restriction on $n$ is a polynomial one (see
[\pref{Lang}].)

It is important for the following to assemble the periods into a {\it
period polynomial}, which, as usual, is a handier quantity,
  $$\eqalign{
  r_\caF(\ssx)&=\sum_{n=0}^{2t-2}(-1)^n\comb{2t-2}{n}\,r_n(\caF)\,
  \ssx^{2t-2-n} \cr
&=\int_0^{\infty}d\sssb\,(\ssx-\sssb)^{2t-2}\caF(i\sssb)
 \,.}
  \eql{ppol7}$$

In order to extend the notion to {\it non}--cusp forms (such as
Eisenstein series) one course (see Zagier, [\pref{Zagier}] and Kohnen and
Zagier, [\pref{KandZ}] \S4), is to continue the integer, $n$, into the
complex plane by employing the Hecke $L$--series of $\caF$ via the Mellin
transform,
  $$
  L(\caF,s)={(2\pi)^s\over\Ga(s)}\int_0^{\infty}dt\,\big(\caF(it)
 -\caF(i\infty)\big)\,t^{s-1}\,,
  \eql{ell}$$
which converges for $s>2t-2$ and, continued in $s$, satisfies the
functional relation,
  $$
  (2\pi)^{-s}\,\Ga(s)\,L(\caF,s)=(-1)^t(2\pi)^{s-2t}\,\Ga(2t-s)\,L(\caF,2t-s)\,,
  \eql{lfr}$$
by virtue of the (inversion) modularity of $\caF$, and conversely. This
is usually attributed to Hecke, [\pref{Hecke}] (but see the Appendix).
Terras, [\pref{Terras}] p.229, gives a useful pedagogical treatment of
this topic.

Then the periods could be {\it defined}, in general, by,
[\pref{KandZ}],[\pref{Zagier}],
  $$\eqalign{
  r_n(\caF)&=\lim_{s\to n+1}\,{\Ga(s)\over(2\pi)^{s}}\,L(\caF,s)
  \cr
  &\equiv\lim_{s\to n+1}L^*(\caF,s)\,,
  }
  \eql{periods2}$$
because this coincides with the integral expression, (\peq{periods}), for
cusp forms.

The explicit values of $r_n$ ($n=0,\ldots 2t-2$) for Eisenstein series
(which is all that is required because of a decomposition theorem for
modular form space) are easily computed from (\peq{eisl1}) and are given
in [\pref{KandZ}]. They agree with (\peq{even}) and (\peq{moments}), as
already noted.

However, the Dirichlet series, $L^*(\caF,s)$ has poles at $s=0$ and
$s=2t$. A means of incorporating these values is given by Zagier,
[\pref{Zagier}], using the period polynomial. One notes that the
summation in the cusp period polynomials can be extended,
  $$\eqalign{
  r_\caF(\ssx)&=\sum_{n\in\,\oZ}(-1)^n\comb{2t-2}{n}\,r_n(\caF)\,
  \ssx^{2t-2-n} \cr
&=\sum_{n\in\,\oZ}(-1)^n\comb{2t-2}{n}\,i^{n+1}\,L^*(\caF,n+1)\,
\ssx^{2t-2-n}\,,}
  \eql{ppol6}$$
because the binomial coefficient vanishes outside the range
$n=0,\ldots,2t-2$. For {\it non}--cusp forms, however, the sum extends
from $-1$ to $n=2t-1$, the nonzero end values arising from the
singularities of $L^*(\caF,s)$ mentioned above. The resulting expression
can then be taken as a {\it definition} of the period polynomial of a
non--cusp form.

Again, for the Eisenstein series, $G_{2t},\,=\ep_{t}$, the computation is
straightforward, [\pref{Zagier}], and (\peq{ppol6}) with (\peq{eisl1})
yields, ($n=2j-1$),
  $$\eqalign{
  r_G(x)={(2t-2)!\over2(2\pi)^{2t}}\bigg(2\pi
  \ze_R&(2t-1)\,\big((-1)^{t-1}\ssx^{2t-2}-1\big)-\cr
  &(2\pi)^t\sum_{j=0}^t\,(-1)^{t-j}{B_{2j}\,B_{2t-2j}\over(2j)!(2t-2j)!}
  \,\ssx^{2j-1}\bigg)\,,
  }
  \eql{eperiodp}
  $$
which is $1/\ssx$ times a polynomial. For comparison with
[\pref{Zagier}], Zagier's $X$ equals my $i\ssx$.

I now comment on this construction. The periods, and period polynomials,
for non--cusp forms are defined by analogy to those of cusp forms, the
justification being that the expressions, for $\caF(i\infty)$ equal to
zero, reduce to those for cusp forms. The means to attain this is not
unique. Instead of the `zeta--function' regularisation, as in the
definition of $L$, (\peq{ell}), I employ the {\it subtracted} form, as in
(\peq{rege}),
  $$
   \caF^{\rm sub}\equiv \caF-a_0-{a_0\over\tau^{2t}}\,,\quad
   a_0=\caF(i\infty)\,,
  $$
whose moments are non--infinite and provide an alternative definition of
the periods of a non--cusp form which equally reduces to that for cusp
forms when $a_0=0$. This subtraction maintains inversion modularity, but
violates translational, whereas the subtraction of just $a_0$ does the
reverse. Making use of my earlier considerations of Eisenstein series,
see (\peq{ppol}), I define the subtracted period polynomial as an
integral,
  $$
  r_\caF^{\rm sub}(\ssx)=\int_0^{\infty}d\sssb\,
  (\sssb-\ssx)^{2t-2}\caF^{\rm sub}(\sssb)\,.
  $$

For Eisenstein series, (\peq{ppol}) shows that $r_G^{\rm sub}(\ssx)$ is
proportional to $\overline P_t(\ssx)$ whose explicit form is given in
(\peq{ppoly2}).

It is necessary to compare this definition with that of Zagier,
(\peq{eperiodp}). The difference is the absence of the `end point' terms
proportional to $1/\ssx$ and $\ssx^{2t-1}$. This can be remedied by
including, in the terminology of this paper, the regularised
contributions from the subtracted terms, the Casimir and Planck terms,
which can be accomplished simply by expanding the closed contour in
(\peq{ppoly2}) to include the corresponding poles at $s=0$ and $s=2t$.
This just extends the sum by the two end points giving the well known
quantity,
   $$\eqalign{
\overline R_t(\ssx)=2\pi\,\ze_R(2t-1)\big((i \ssx)^{2t-2}-1\big)-4i
  \sum_{j=0}^t\ze_R(2j)\,\ze_R(2t-2j)(i \ssx)^{2t-2j-1}
  \,,}
  \eql{ppoly3}
  $$
which agrees, up to a factor, with Zagier's expression given in
(\peq{eperiodp}).

Furthermore, the use of the subtracted form, $\caF^{\rm sub}$, allows one
to give a cleaner integral expression for the enlarged period polynomial
than that in [\pref{Zagier}], \ie,
  $$
  r_\caF^{\rm enl}(\ssx)=\int_0^{\infty}d\sssb\,
  (\sssb-\ssx)^{2t-2}\caF^{\rm sub}(\sssb)+
  {a_0\over2t-1}\big(\ssx^{2t-1}+\ssx^{-1}\big)\,.
  $$

While equivalent to the method in [\pref{Zagier}], I believe the contour
approach is neater and is capable of being extended to the general theory
of periods. Zagier, [\pref{Zagier}], shows that logistic simplifications
occur in this theory when it is extended to include non--cusp forms such
as the Eisenstein series.

As an historical point, it will be noticed that the above expressions
have been available since the time of Ramanujan, with contour derivations
by Malurkar, [\pref{Malurkar}], and Guinand, [\pref{Guinand}]. In order
to investigate this point further, I return to the mode sum expressions
for the thermodynamic related quantities.

\section{ \bf 3. Back to summations.}
I start with the Eisenstein series forms and work the development to
parallel the quantities of the previous section. This allows me to
introduce the useful calculations of Smart, [\pref{Smart}], on the
Epstein function (which will arise later) using its relation with
Eisenstein series.

The subtracted partial energy $\ep_t^{\rm sub}$ in (\peq{rege})
corresponds to the the `regularised' Eisenstein series,
  $$\eqalign{
  G_t^{\,\rm sub}(\tau)&=\sumdasht{m=-\infty}{\infty}\,\sumdasht{n=-\infty}{\infty}
  {1\over (m\tau+n)^{2t}}\cr
  &=\,G_t(\tau)-2\ze_R(2t)(1+\tau^{-2t})\,,}
  \eql{geesub}$$
The relation is, [\pref{DandK1}],
  $$
 \ep^{\rm sub}_t(\tau)={(2t-1)!\over2(2\pi i)^{2t}}\, \overline
 G^{\,\rm sub}_t(\tau)\,.
  \eql{inten5}$$

Now, corresponding to (\peq{mint2}), integrate (\peq{geesub}) with
respect to $\tau$ from $\tau$ to $i\infty$, $2t-1$ times to give,
formally,
  $$
  {1\over(2t-1)!}\,\phi_{2t}(\tau)\,,
  \eql{geeint}$$
where $\phi$ is defined by, [\pref{Smart}], eqn.(2.11),
  $$
  \phi_{2t}(\tau)=\sumdasht{m=-\infty}{\infty}\,\sumdasht{n=-\infty}{\infty}
  {1\over m^{2t-1}}{1\over m\tau+n}\,.
  \eql{phis}$$

The behaviour of the series $\phi_{2k}(\tau)$ under modular
transformations, in particular $\tau\to-1/\tau$, has been especially
considered by Smart, [\pref{Smart}], who refers to the $\phi$'s as `{\it
modular forms of weight $2-2t$ with rational period functions}'.

Using the partial fraction identity (see also Hurwitz, [\pref{Hurwitz}],
Eisenstein, [\pref{Eisenstein}]),
  $$
     {1\over m^{2t-1}}{1\over m\tau+n}=
     -{\tau^{2t-1}\over n^{2t-1}}{1\over m\tau+n}+\sum_{j=1}^{2t-1}(-1)^{j+1}
     {\tau^{2t-j-1}\over m^j\,n^{2t-j}}\,,
  \eql{pf}$$
one finds easily the inversion relation, [\pref{Smart}], eqn.(1.10b),
  $$
  \phi_{2t}(\tau)-\tau^{2t-2}\phi_{2t}(-1/\tau)=-4\sum_{j=1}^{t-1}
     \tau^{2t-2j-1}\ze_R(2j)\,\ze_R(2t-2j)\,.
  \eql{sm1}$$
Also, under translations,
  $$
  \phi_{2t}(\tau+1)-\phi_{2t}(\tau)=2{\ze_R(2t)\over\tau(\tau+1)}\,.
  \eql{sm2}$$

Up to a simple factor, one might expect $\phi_{2t}$ to be the same as
${\overline\phi}_{2t}$ in (\peq{mint2}). However the cocycle functions in
(\peq{sm1}) and (\peq{ppoly2}) differ by the two {\it even} powers. I
note that these are related to the $j=1$ and $j=2t-1$ terms in the sum in
(\peq{pf}) which have gone out in the passage to the difference,
(\peq{sm1}), by antisymmetry. However the summation over $m$,
respectively $n$, for these values of $j$ is conditionally convergent and
is therefore subject to ambiguity, that is to say, a choice has to be
made, as in the discussion of the Eisenstein series, $G_2$. The
regularised Mellin transform approach results in a different definition
of the series (\peq{phis}) to that used by Smart, [\pref{Smart}],p.3. The
relation is easily found by taking the even powers in (\peq{ppoly2}) over
to the left and then we see that one has the relation, (further remarks
are given later),
  $$
  {\overline\phi}_{2t}(\ssx)=
i\phi_{2t}(\tau)-2\pi \ze_R(2t-1)\,,\quad \ssx=-i\tau\,,
  \eql{connec}
  $$
also agreeing with (\peq{trans}) and (\peq{sm2}).

Smart, [\pref{Smart}], also gives the relation with a known Lambert
series as
  $$\eqalign{
  S_t(\tau)\equiv\sum_{m=1}^\infty{1\over m^{2t-1}}{q^{2m}\over1-q^{2m}}
  &=-{1\over4\pi i}\,\phi_{2t}(\tau)-{1\over2\pi i\tau}\,
\ze_R(2t)-{1\over2}  \ze_R(2t-1)\cr
&={1\over4\pi}\,\big(\overline\phi_{2t}(\ssx)+{2\over\ssx}\ze_R(2t)\big)\cr
&\equiv{1\over4\pi}\,\overline\psi_{2t}(\ssx)\,,
}
  \eql{ls}$$
where $\overline\psi_{2t}$ is the function having $\overline R_{2t}(\ssx)$,
(\peq{ppoly3}), as its inversion cocycle function,
  $$
  {\overline\psi}_{2t}(\ssx)-(i\ssx)^{2t-2}\,{\overline\psi}_{2t}(1/\ssx)
= \overline R_{2t}(\ssx)\,,
  \eql{are1}$$
as is easily confirmed.

The introduction of $\overline\psi$ has restored translational
invariance, (the Planck term has been put back in (\peq{ls})),
  $$
  \overline\psi_{2t}(\ssx-i)- \overline\psi_{2t}(\ssx)=0\,.
  \eql{per}$$
It is almost convention to consider modular integrals, for example, to be
periodic (\eg Knopp, [\pref{Knopp}]). See some later comments.

Equations (\peq{are1}) and (\peq{ls})  imply an inversion relation for
the series $S_t$ which has a certain history, some of which is detailed
by Berndt, [\pref{Berndt}], p.153, who says that it was first written
down by Ramanujan. As is clear, it is quite the same as the expression
for the Eisenstein period polynomials.

The Lambert series, $S_t(\tau)$, has been considered by Apostol,
[\pref{Apostol}], Berndt, [\pref{Berndt}], Grosswald, [\pref{Grosswald}],
and Smart [\pref{Smart}], for example. It can be related to statistical
mechanics in the following way.

One can resum the statistical expression for the partial free energy in a
standard manner, beginning, ($q=e^{\pi i\tau}=e^{-\pi/\xi}$),
  $$
  f_t=\ep_{t,0}-{\xi\over2\pi}\sum_{n=1}^\infty n^{2t-2}\log(1-q^{2n})
  =\ep_{t,0}-{\xi\over2\pi}\sum_{m,n=1}^\infty{n^{2t-2}\over m}q^{2mn}\,.
  \eql{ssum}$$

The scaled entropy is defined by,
  $$
  s_t(\xi)={\pa f_t(\xi)\over\pa \xi}=-{1\over2\pi}\bigg(1-{2\pi\over\xi}D
\bigg) \sum_{m,n=1}^\infty{n^{2t-2}\over m}q^{2mn}\,.
  \eql{scent}$$

Concentrate now on the summation and rewrite it,
  $$
  \sum_{m,n=1}^\infty{n^{2t-2}\over m}q^{2mn}=D^{2t-2}
\sum_{m=1}^\infty{1\over m^{2t-1}}{q^{2m}\over1-q^{2m}}\,,
  \eql{sum9}$$
in terms of $S_t(\tau)$, (\peq{ls}).

This Lambert series  has been discussed in connection with generalised
Dedekind $\eta$--functions by Berndt, [\pref{Berndt}], and it is worth
noting that, in a certain sense, $\log \eta$ is a modular form of zero
weight, in agreement with the results $\ep_1=-D\log\eta$,
[\pref{DandK2}], which could therefore be written,
  $
  \ep_1=-\caD\log\eta\,,
  $
in terms of the covariant derivative, $\caD$.

The question, not answered here, is whether there is an `expansion' or
representation theorem for modular {\it integrals} combining that for
modular forms (of {\it any} real weight, see [\pref{Rankin}]). It is not
likely that one can find a pure elliptic formula. As evidence I cite the
particular evaluations of Smart who gives at the lemniscate point
($\tau=i,\ssx=1$),
  $$
  \overline\psi_{4}(1)={7\pi^4\over90}-2\pi\ze_R(3)\,,
  $$
and, more generally, if $2t=0({\rm mod} 4)$,
  $$
  S_t(i)={\ze_R(2t)\over2\pi}-{\ze_R(2t-1)\over2}+{1\over2\pi}
  \sum_{j=1}^{t-1}(-1)^{j+1}\ze_R(2t-2j)\ze_R(2j)\,,
  $$
a result actually due to Lerch which follows from (\peq{are1}) with
(\peq{ppoly3}).

As a final general comment in this section, it is always possible to
obtain the required quantities as (infinite) $q$-series by direct
integration, but these do not count as closed forms. For example, one
integration yields the $q$-series for the partial free energy in terms of
the arithmetic functions, $\si_k(m)$, (the sum of the $k$-th powers of
the divisors of $m$),
  $$
  f_t=\ep_{t,0}+{1\over\be}\sum_{m=1}^\infty\si_{2t-1}(m)\,{q^{2m}\over m}\,,
  $$
which amounts only to doing the $n$-summation in ({\peq{ssum}), or an
integration of the standard relation, (\peq{mell1}),
  $$
  \ep_t=-{B_{2t}\over4t}+\sum_{m=1}^\infty \si_{2t-1}(m)\,q^{2m}\,.
  $$

One also has the formula, \eg [\pref{Rad,Glaisher}],
  $$
  S_t(\tau)=\sum_{m=1}^\infty \si_{-2t+1}(m)\,q^{2m}=
  \sum_{m=1}^\infty {\si_{2t-1}(m)\over m^{2t-1}}\,q^{2m}\,,\quad\forall\, t\,,
  \eql{Ssi}$$
in agreement with the result of a $(2t-1)$-fold integration.
\section{\bf 4. Another approach via modular integrals.}
The period functions are cocycle functions associated with Eichler
cohomology and modular integrals. These notions put a different slant on
the preceding formulae. In effect one begins again.

The best place to start is the important (but particular) analytical result
of Bol, [\pref{Bol}], which states that, for {\it any} reasonable function,
$\varphi(\tau)$,
  $$
  \big(D^{r+1}\varphi\big)(\ga\tau)=
  (c\tau+d)^{r+2}D^{r+1}\big((c\tau+d)^r\,\varphi(\ga\tau)\big)\,,
  \eql{bol}$$
where $r$ is a nonnegative number, and I use,
  $$(D^{r+1}\varphi)(\tau)\equiv \bigg(q{d\over dq}\bigg)^{r+1}\varphi(\tau)\,,
  $$
with,
  $$
   \ga=\left(\matrix{*&*\cr
                   c&d}\right)\in {\rm SL}(2,\oR)\,.
  $$
This shows that, {\it if} $\varphi$ is an automorphic form of weight
$-r$, then $D^{r+1}\,\varphi$ is an automorphic form of weight $2+r$, see
also Petersson, [\pref{Petersson}]. Note that it is the ordinary
derivative that enters here. The proof of (\peq{bol}) follows by
induction.

The classic discussion of {\it modular integrals} can proceed by asking
for the inverse of Bol's result. That is, given a form of weight $2+r$,
$\rho(\tau)$, as a source, can one integrate the differential equation,
  $$
  D^{r+1}\varphi=\rho
  \eql{de}$$
to obtain $\varphi$, a form of weight $-r$? Clearly this will not be so
in general because of the freedom introduced by constants of integration.
That is, one would expect $\varphi$ to behave as a form of weight $-r$ up
to an additional piece, the general form of which can be determined from
the differential equation (\peq{de}) and Bol's theorem, (\peq{bol}).

Generally we would write (ignoring any multiplier systems for
simplicity),
  $$
  \varphi(\ga\tau)=(c\tau+d)^{-r}\big(\varphi(\tau)+P(\ga,\tau)\big)\,,
  \eql{co2}$$
where $P(\ga,\tau)$ has to be a polynomial in $\tau$ of order $\le r$,
The inverse to Bol's result is that, if any $\varphi$ satisfies
(\peq{co2}), then $\rho$, of (\peq{de}), is a form of weight $(2+r)$.
This follows most simply by substituting (\peq{co2}) into (\peq{bol}).

$P$ must satisfy the 1--cocycle consistency condition, [\pref{Eichler}],
  $$\eqalign{
   P(\ga_1\ga_2,\tau)&=(c_2\tau+d_2)^r\,P(\ga_1,\ga_2\tau)+P(\ga_
   2,\tau)\cr
   &\equiv P(\ga_1,\tau)\big|\ga_2+P(\ga_2,\tau)\,.
   }
  \eql{cocycle}$$

$\varphi$ is known as an {\it automorphic} or {\it Eichler integral}, of
weight $-r$, and $P$ is the {\it associated period polynomial},
[\pref{Eichler,HandK}]. I have also introduced the useful, standard
`stroke' operator, $\big|\ga$, sometimes denoted $\big|[\ga]$.

The {\it general} solution (equivalently the indefinite integral) of the
differential equation (\peq{de}) is a particular integral plus the
complementary function (a zero mode) \ie
  $$
  \varphi(\tau)={(2\pi i)^{r+1}\over\Ga(r+1)}
  \int_{\tau_0}^\tau d\si\, (\tau-\si)^{r}\,\rho(\si)+\Th(\tau)\,,
  \eql{gensol}
  $$
where $\Th(\tau)$ is a polynomial of degree $r$. The integral, in the
upper half plane, is path independent if $\rho$ has weakish analytic
properties and, if desired, the solution can be made independent of the
lower limit by suitable adjustment of $\Th(\tau)$. In elementary fashion,
$\Th(\tau)$ is determined by the first $r$ derivatives of $\varphi(\tau)$
at $\tau_0$.

All this is general analysis. In modular applications to SL(2,$\oZ$),
$\tau_0$ is chosen as the cusp $i\infty$ and the limits in (\peq{gensol})
reversed. In terms of Fourier expansions, a holomorphic cusp form source
is
  $$
  \rho(\tau)=\sum_{m=1}^\infty \rho_m\,q^{2m}\,,
  \eql{fe1}$$
and the particular Eichler integral of $\rho$, is, from (\peq{gensol}),
  $$
  \varphi(\tau)=\sum_{m=1}^\infty {\rho_m\over m^{r+1}}\,q^{2m}\,,
  \eql{fe2}$$
which has the same periodicity as $\rho$, a result of choosing
$\tau_0=i\infty$.

The problem in the present Eisenstein case has already been noted. The
Eisenstein series, $G_t(\tau)$ does not vanish at $i\infty$ (or $0$). It
is a `non--cusp' form and the direct integration of (\peq{de}) meets
obstacles at an elementary level. This can be appreciated simply from the
appearance of $1/m$ factors during integration, and $m$ can be zero in
$G_t$. This fits in because the $m=0$ terms in $G_t$ are those making
$G_t$ nonzero at $i\infty$, the zero temperature value. Saying it yet
again, the Fourier transform of $G_t$ begins with an $m=0$ term, the
Casimir contribution, so (\peq{fe2}) makes no immediate sense.

One way of avoiding these obstructions, is to regularise $G_t$ by
dropping the offending term(s) as in (\peq{geesub}) and consider the
differential equation,
  $$\eqalign{
  D^{2t-1}\,\varphi_{2t}=-4\pi\ep_t^{\rm sub}\,,}
  \eql{gee}$$
which has the particular integral, $\varphi_{2t}=\overline\phi_{2t}$
where,
  $$
  \overline\phi_{2t}(\tau)=-4\pi{(2\pi i)^{2t-1}\over\Ga(2t-1)}
  \int_\tau^{i\infty}d\tau'\,(\tau'-\tau)^{2t-2}
  \,\ep_t^{\rm sub}(\tau')\,,
  $$
the constants and boundary conditions being chosen so as to reproduce
(\peq{mint2}). The upper limit is a cusp. Smart's function, $\phi_{2t}$,
results if one adds a complementary constant (and multiplies by $i$).
(See (\peq{connec}).)

Because of the subtraction, the quantity $\ep_t^{\rm sub}(\tau)$ is not
exactly a modular form, rather one has, expressed slightly generally,
($r=2t-2$),
  $$
  \ep(\ga\tau)=(c\tau+d)^{r+2}\big(\ep(\tau)+E(\ga,\tau)\big)
  \eql{submod}
  $$
and also
  $$
  \phi(\ga\tau)=(c\tau+d)^{-r}\big(\phi(\tau)+P(\ga,\tau)\big)\,,
  \eql{co3}
  $$
which are related by the same differential equation, (\peq{de}),
  $$
  D^{r+1}\phi=-4\pi\ep\,.
  $$
Bol's formula, (\peq{bol}),  relates $P$ and $E$,
  $$
  D^{r+1}P=-4\pi E\,,
  \eql{bol2}
  $$
and these relations generalise the usual ones applicable to cusp forms.
The cocycle formula (\peq{cocycle}) still holds of course, as (\peq{co3})
is the same as (\peq{co2}).

I remark that $\overline\phi$ and $\ep^{\rm sub}$, not being periodic, do
not quite possess Fourier expansions, as in (\peq{fe1}) and (\peq{fe2}).
However $\overline\psi$, in view of (\peq{per}), does,
  $$
  \overline\psi(\ssx)=\sum_{m=1}^\infty \psi_m\,q^{2m}\,,
  $$

Because $P$, this time, satisfies (\peq{bol2}), it is not required to be
a polynomial. Generally, since the modular group is generated \footnote{
Beware that some authors, \eg Knopp, swap the usage of the symbols $S$
and $T$.} by $S$ and $T$, $P(\ga,\tau)$ is a {\it rational period
function}. In the present case, however, $P(S,\tau)$ {\it is} a
polynomial, given by (\peq{ppoly2}),
  $$
  P(S,\tau)=\overline P_t(\tau)\,,
  \eql{pt}$$
while the (non--zero) $P(T,\tau)$, which follows from (\peq{trans}),
  $$
  P(T,\tau)=2\,\ze_R(2t)\bigg({1\over \tau}-{1\over\tau+1}\bigg)\,,
  \eql{ps}$$
is a ratio of polynomials. Of course, $P(\pm{\bf 1},\tau)=0$ and I remark
that the polynomial $P(S,\tau)$ runs from $\tau^0$ to $\tau^{2t-2}$.

As a simple check of the algebra, set $\ga_1=\ga_2=S$ in the cocycle
relation, (\peq{cocycle}). This gives,
  $$
 P(S,\tau)\big|(1+S)\equiv P(S,\tau)+\tau^{r}P(S,S\tau)=
 P(S,\tau)+\tau^{r}P(S,-1/\tau)=0
  $$
which, for $r=2t-2$, agrees with the explicit form, (\peq{ppoly2}).

Also, as examples,
  $$\eqalign{
  P(TS,\tau)=&2\,\ze_R(2t){\tau^{2t}\over1-\tau}+\ol P_t(\tau)\cr
  P(ST,\tau)=&{2\,\ze_R(2t)\over\tau(1+\tau)}+\ol P_t(1+\tau)\,.}
  $$
I note that the period function appears to have poles at $-1$, $0$, $+1$
and $i\infty$.

From (\peq{submod}) and the form of $\ep^{\rm sub}$, (\peq{rege}), one
finds
  $$
  E(S,\tau)=0\,,\quad E(T,\tau)=\bigg({1\over(\tau+1)^{2t}}
  -{1\over\tau^{2t}}\bigg)
  $$
which are consistent with (\peq{ps}) and (\peq{pt}).

For completeness, and to aid comparison, this is a convenient place to
outline some standard elementary things about period polynomials, mostly
for SL$(2,\oZ)$. Two basic `polynomials' are the inversion one,
$P(S,\tau)$, and the translation one, $P(T,\tau)$. For cusp forms, the
latter vanishes because of periodicity, $P(T,\tau)=0$, [\pref{Knopp}],
corresponding to the periodicity of the modular integrals, \cf
(\peq{fe1}), (\peq{fe2}). In this case, $P(S,\tau)$ satisfies the
Eichler--Shimura relations
  $$\eqalign{
  P(S,\tau)\big|(1+S)&=0 \cr
  P(S,\tau)\big|(1+TS+(TS)^2)&=0
  }
  \eql{ES}
  $$
with $S^2=1$ and $(TS)^3=1$. A standard way of proving these uses the
complex integral form for $P$, \eg [\pref{Zagier}], but they  can also
follow, algebraically from the cocycle relation as I now show. (See also
Knopp, [\pref{Knopp3}], for slightly different details.)

For the first identity, just set, as above, $\ga_1=1$ and $\ga_2=S$ in
(\peq{cocycle}) and it falls straight out. The second identity is
slightly more work. Using (\peq{cocycle}) twice, one has, dropping the
$\tau$ argument temporarily and noting $(TS)^2=(TS)^{-1}$,
  $$\eqalign{
  P(S)\big|\big(1+&TS+(TS)^2\big)=\cr
  &P(S)+P(STS)-P(TS)+P(SS^{-1}T^{-1})-P(TSTS)\,.
  }
  \eql{ES2}
  $$

Further, setting $\ga_1=T$ in the cocycle condition, and using the
assumption that $P(T)=0$, yields
  $$
  P(T\ga)=P(\ga)\,.
  \eql{pt0}
  $$
One then sees that the first and third terms on the right--hand side of
(\peq{ES2}) cancel, as do the second and fifth. The fourth term vanishes
on its own because (\peq{pt0}) shows that $P(T^{-1})=0$, and (\peq{ES})
follows.

In one development, the notion of modular integral has been somewhat
divorced from that of integration. One says that a {\it modular integral}
is any function (with appropriate analyticity properties) that obeys the
quasi--modular relation (\peq{co2}), where the {\it period function} $P$
must still, of course, satisfy the cocycle condition (\peq{cocycle}).
Motivation is then provided to look for modular integrals with {\it
rational} period functions, \ie ratios of polynomials. In this case it
seems to be conventional to {\it assume} that the translational period
function, $P(T,\tau)$, vanishes, [\pref{Knopp}], so that both
Eichler--Shimura relations, ({\peq{ES}), still hold. There is a certain
interest in computing such rational period functions and investigating
their zeros and poles. The easiest introduction to this topic is by
Knopp, [\pref{Knopp3}]. It is, however, somewhat peripheral to our
considerations which are more concerned with non--cusp forms \ie the
Eisenstein series, \eg\ [\pref{Zagier}].

\section{\bf 5. Green function distributional description.}

Developing the formalism a little further, it is possible to give the
discussion of Eichler integrals a distributional or Green function look.
For this purpose, it is better to use the `real' variable form and consider
the differential equation,
  $$
  {d^{2t-1}\over d{\ssx}^{2t-1}}\,\psi(\ssx)=\rho(\ssx)\,,
  \eql{rde}$$
for $\ssx>0$ with solution vanishing at $\infty$, \cf (\peq{mint2}),
  $$
  \psi(\ssx)=
\int_{0^-}^\infty d{\ssx'}\,\Phi_{2t-1}(\ssx'-\ssx)\,\rho({\ssx'})
  \eql{soln}$$
where the generalised function, $\Phi_\al(\ssx)$, is
 $$
\Phi_\al(\ssx)={\ssx^{\al-1}_+\over\Ga(\al)}\,.
 \eql{genp}$$
The {\it generalised} function $\ssx^\al_+$ is concentrated on the positive
$\ssx$-axis, \ie $\ssx^\al_+$ is equal to $\ssx^\al$ for $\ssx\ge0$ and is
zero for $\ssx<0$, \eg Gelfand and Shilov, [\pref{GandS}] I,\S5.5.
$\Phi_{2t-1}$ acts as a Green function satisfying,
  $$
  {d^{2t-1}\over d{\ssx}^{2t-1}}\,\Phi_{2t-1}(\ssx'-\ssx)=\de(\ssx'-\ssx)\,.
  $$
The formal, distributional equivalent is obtained from the convolution, valid
for all $\al$ and $\be$,
  $$
  \Phi_\al*\Phi_\be=\Phi_{\al+\be}\,,
 \eql{conv}$$
by setting $\al=-\be=2t-1$ and noting
  $$
\Phi_{-k}(\ssx)=\de^{(k)}(\ssx)\,,\quad k=0,1,2,\ldots.
 \eql{deriv}$$

An important formula is the Fourier transform of $\Phi$, [\pref{GandS}]
I.p.360,
  $$
  \int_0^\infty d\ssx\, \Phi_\al(\ssx)\,e^{i\si \ssx}
  ={1\over (\si+i0)^\al}\,e^{i\pi\al/2}\,,
  \eql{ftphi}$$
and the inverse, which is a (continuous) eigenfunction expansion,
  $$
  \Phi_\al(\ssx)=e^{i\pi\al/2}{1\over2\pi}\int_{-\infty}^\infty d\si\,
  {e^{-i\si \ssx}\over(\si+i0)^\al}\,.
  \eql{invft}$$

In the present case, the source $\rho$ is assumed to be periodic  under
$\ssx\to\ssx-i$ (the Planck term is not removed) and to converge as
$\ssx\to\infty$. So one has the `Fourier' series,
  $$
  \rho(\ssx)=\sum_{m=1}^\infty \rho_m\, e^{-2m\pi\ssx}\,,\quad \ssx>0\,.
  \eql{ftrho}$$
Substitution of this and (\peq{invft}) into (\peq{soln}) gives,
  $$
  \psi(\ssx)=(-1)^{t+1}{1\over2\pi}\sum_{m=1}^\infty\rho_m
 \int_{-\infty}^\infty d\si\,
  {e^{i\si\ssx}\over(\si+i0)^{2t-1}(\si-2m\pi i)}\,.
  \eql{psi1}
  $$
Closing the contour in the upper half $\si$--plane I obtain, $\ssx>0$,
  $$
  \psi(\ssx)={1\over(2\pi)^{2t-1}}\sum_{m=1}^\infty
  {\rho_m\over m^{2t-1}}\,e^{-2m\pi\ssx}\,,
  \eql{soln2}$$
which, it is no surprise to see, is just (\peq{fe2}) derived by a much
longer route but one on which we encounter some handy information. It is
seen also that the source periodicity has been transmitted to the solution.

Including an $m=0$ term in the Fourier series, (\peq{ftrho}), still gives
convergence at infinity, but of course the solution, (\peq{soln2}),
breaks down, as has already been mentioned. As with all Green function
approaches, when this happens the corresponding eigenfunction has to be
removed from the source. This can be done formally by redefining the
Green function to exclude this function which is, in the present case, a
constant zero mode. Altering the contour in the Fourier integral
(\peq{invft}) gives a modified Green function,
  $$
  \Phi^{\rm sub}_{2t-1}(\ssx)\equiv(-1)^{t}{1\over2\pi i}
  \int_{-\infty+i\de}^{\infty+i\de} d\si\,
  {e^{-i\si\ssx}\over\si^{2t-1}}\,,\quad 0<\de<2\,,
  \eql{phisub}$$
which automatically takes out any constant part of the source. This can
be seen by going back to (\peq{psi1}) and replacing the integral by one
from $-\infty+i\de$ to $\infty+i\de$,
  $$
  \psi(\ssx)=(-1)^{t+1}{1\over2\pi}\sum_{m=1}^\infty\rho_m
 \int_{-\infty+i\de}^{\infty+i\de} d\si\,
  {e^{i\si\ssx}\over\si^{2t-1}(\si-2m\pi i)}\,.
  \eql{psi2}
  $$
For $ 0<\de<2$, if $\ssx>0$, closing the contour in the upper half plane
still yields (\peq{soln2}). However now, even if the source Fourier
series, (\peq{ftrho}), is extended to the constant term, $m=0$, the
result is again (\peq{soln2}). This should be denoted by $\psi^{\rm
sub}(\ssx)$ and one can write
  $$
  \psi^{\rm sub}(\ssx)=
\int_{0^-}^\infty d{\ssx'}\,\Phi^{\rm
sub}_{2t-1}(\ssx'-\ssx)\,\rho({\ssx'})\,.
  \eql{soln3}$$

Retaining the general structure, (\peq{invft}), for any $\al$, one can
define for any $\al$,
  $$
  \Phi^{\rm sub}_\al(\ssx)\equiv e^{i\pi\al/2}{1\over2\pi}
\int_{-\infty+i\de}^{\infty+i\de}d\si\,
  {e^{-i\si \ssx}\over\si^\al}\,.
  \eql{invft2}$$
Then $\Phi^{\rm sub}_\al(\ssx)$ is still concentrated on the positive
$\ssx$--axis and the convolution (\peq{conv}) also remains valid,
  $$
  \Phi^{\rm sub}_\al* \Phi^{\rm sub}_\be =\Phi^{\rm sub}_{\al+\be}
  \eql{conv2}
  $$
together with (\peq{deriv}),
  $$
  \Phi^{\rm sub}_{-k}=\de^{(k)}\,.
  \eql{deriv2}
  $$

To check these statements, firstly, if $\ssx<0$, closing the contour in
(\peq{invft2}) in the upper half plane yields zero as the only
singularity is at $\si=0$. Then
  $$\eqalign{
  \big(\Phi^{\rm sub}_\al* \Phi^{\rm sub}_\be\big) (\ssx)=
 {e^{i\pi(\al+\be)/2}\over(2\pi)^2}\int_{-\infty}^\infty d\ssx'
\int_{-\infty+i\de}^{\infty+i\de}d\si\,\int_{-\infty+i\de}^{\infty+i\de}d\si'\,
  {e^{-i\si\ssx-i\ssx'(\si'-\si)}\over\si^\al\si'^\be}
  }
  $$
where the lower limit on the $\ssx'$ integral is allowed because
$\Phi_\be^{\rm sub}(\ssx')$ is concentrated on the positive $\ssx'$ axis.

Setting $\si=s+i\de$, $\si'=s'+i\de$, I get, in detail,
  $$\eqalign{
 \big(\Phi^{\rm sub}_\al* \Phi^{\rm sub}_\be\big) (\ssx)&= e^{i\pi(\al+\be)/2}{1\over(2\pi)^2}\int_{-\infty}^\infty d\ssx'
\int_{-\infty}^{\infty}ds\,\int_{-\infty}^{\infty}ds'\,
  {e^{-i\si\ssx-i\ssx'(s'-s)}\over\si^\al\si'^\be}\cr
  &= e^{i\pi(\al+\be)/2}{1\over2\pi}
\int_{-\infty}^{\infty}ds\,\int_{-\infty}^{\infty}ds'\,
  {e^{-i\si\ssx}\over\si^\al\si'^\be}\,\de(s-s')\cr
   &=e^{i\pi(\al+\be)/2}{1\over2\pi}
\int_{-\infty}^{\infty}ds\,
  {e^{-i\si\ssx}\over\si^\al\si^\be}=e^{i\pi(\al+\be)/2}{1\over2\pi}
\int_{-\infty+i\de}^{\infty+i\de}d\si\,
  {e^{-i\si\ssx}\over\si^{\al+\be}}\cr
  &=\Phi^{\rm sub}_{\al+\be}(\ssx)\,.
  }
  $$

  Equation (\peq{deriv2}) follows either from (\peq{conv2}) or
  (\peq{invft2}) since the integrand has no singularity, and so one can
  set $\de=0$.

\subsection{\it Eichler cohomology.}
As a side remark, the general solution to (\peq{rde}) consists of the
particular integral, (\peq{soln}), plus a solution of the homogeneous
equation, \ie a zero mode, which in this case is a polynomial,
$\Th(\ssx)$, of degree at most $2t-2$. The contribution of this to the
cocycle polynomial, $P$ of (\peq{co2}), constitutes an element of the set
of {\it coboundaries} and a cohomology of polynomials can be set up
(Eichler cohomology). Working with the cohomology classes removes the
constants of integration ambiguity in the solution of (\peq{rde}). That
is, there is a one-to-one correspondence between $\rho$ and an element of
the cohomology group denoted $H^1(\Ga,\Pi_r)$, where $\Pi_r$ is the
vector space of polynomials of degree $\le r=2t-2$ (for cusp forms) and
$\Ga$ is (here) the modular group. The development of these ideas, while
significant generally, is not required here.

\section{\bf 6. Epstein approach to thermal quantities.}

An `earlier' form of the free energy, or effective action, follows from
the thermal \zf, which, in this case is related to the Epstein \zf.
Starting from this I will be able to make contact with the Eisenstein
formulation and derive useful relations. Using the Epstein function is
equivalent to the non-holomorphic Eisenstein series. As a rule the
analysis is harder. In reality one is doing twice the necessary work.

In some ways we are going backwards. The use of holomorphic forms, \eg
(\peq{inten5}), should be considered a simplification available through
working with an explicit square root of the Laplacian.

Again splitting up the degeneracy, one is led to define a `partial'
\zf,
  $$
  \ze_t(s,\be)={i\over2\be}\sumdasht{{m,n=\atop-\infty}}
{\infty}{n^{2t-2}\over
 \big(4\pi^2m^2/\be^2+n^2/a^2\big)^s}\,,
  \eql{epthzf}$$
and the related free energy,
  $$
  \overline F_t=-{\xi\over8\pi}\lim_{s\to0}{1\over s}\,
  \sumdasht{{m,n=\atop-\infty}}{\infty}{n^{2t-2}\over
 \big(m^2+n^2\xi^{-2}\big)^s}\,.
  \eql{eff}$$

I have rescaled the free energy by defining, as above and as in
[\pref{DandK1}], $\overline F_t=a F_t$ and have also taken a preliminary
limit of $s\to0$ in an extracted factor of $(2\pi/\be)^{2s}$. This will
be justified in a moment.
  $$
  \overline F_t=(-1)^{t-1}{\xi\over8\pi}\bigg({d\over d(1/\xi^2)}\bigg)^{t-1}
  \lim_{s\to0}{\Ga(s-t+1)\over\Ga( s+1)}\,
  \sumdasht{{m,n=\atop-\infty}}{\infty}{1\over
 \big(m^2+n^2\xi^{-2}\big)^{s-t+1}}\,.
  \eql{eff2}$$

Not surprisingly, the degeneracy factor is replaced by a derivative and the
resulting sum is precisely an Epstein \zf. This has been used before \eg
[\pref{Kennedy}]. The point is that there is no pole at $s=0$, which
justifies the limit. In terms of the Epstein function, $Z_2$,
  $$
  \overline F_t=(-1)^{t-1}{\xi\over8\pi}\bigg({d\over d(1/\xi^2)}\bigg)^{t-1}
  \lim_{s\to0}\,{\Ga(s-t+1)\over\Ga( s+1)}\,Z_2(2s-2t+2,A)\,,
  \eql{eff3}$$
where $A$ is the matrix (the `modulus'),
  $$
  A=\left(\matrix{1&0\cr
             0&\xi^{-2}}\right)\,.
  $$
The idea now is to use the standard functional relation satisfied by
$Z_2(s,A)$, repeated here for convenience,
  $$
  Z_2(2s,A)={\pi^{2s-1}\over\sqrt{\det A}}{\Ga(1-s)\over\Ga(s)}
  \,Z_2(2-2s,A^{-1})\,,
  \eql{freln}$$
to give
  $$
  \overline F_t={\xi\over8\pi}\bigg(-{d\over d(1/\xi^2)}\bigg)^{t-1}\xi
  \lim_{s\to0}\pi^{2s-2t+1}\,{\Ga(t-s)\over\Ga( s+1)}\,Z_2(2t-2s,A^{-1})\,.
  $$
$Z_2(2s,A)$ has a pole at $s=1$ only, so working at $t>1$ to begin with,
one has
  $$
  \overline F_t=\xi{(-1)^{t-1}\Ga(t)\over8\pi^{2t}}
  \bigg({d\over d(1/\xi^2)}\bigg)^{t-1}\xi
  \,\,Z_2(2t,A^{-1})\,.
  \eql{epder}$$
As the simplest case, set $t=2$ giving the three-sphere. Then
   $$
  \overline F_3={\xi\over8\pi^4}{d\over d(1/\xi^2)}\,\xi\,Z_2(4,A^{-1})=
  -{\xi^4\over16\pi^4}{d\over d\xi}\,\xi\,Z_2(4,A^{-1})\,,
  \eql{3free1}$$
agreeing with the result in [\pref{Kennedy}], allowing for the changes in
notation.

This formula can be taken further as shown by Epstein [\pref{Epstein}]
p.633 who gives,
  $$
  Z_2(4, A^{-1})={\pi^4\over45}+\pi\xi^{-3}\ze_R(3)+2\pi\xi^{-3}\chi(q')
  +4\pi^2\xi^{-2}\,D'\,\chi(q')\,,
  \eql{epz2}$$
where $q'=e^{-\pi\xi}$ and
  $$
  \chi(q')=S_2(-1/\tau)=\sum_{m=1}^\infty {q'^{2m}\over m^3(1-q'^{2m})}\,.
  $$

Equation (\peq{3free1}) with (\peq{epz2}) is appropriate for the high
temperature limit since $q'$ is the `inverse' temperature Boltzmann factor.
The low temperature form follows by simple homogeneous scaling of the
Epstein \zf, which gives,
  $$
  \,Z_2(2s,A^{-1})=\xi^{-2s}Z_2(2s,A),
  \eql{scale}$$
and therefore the equivalent form,
  $$
  Z_2(4, A^{-1})={\pi^4\over45}\xi^{-4}+\pi\xi^{-1}\ze_R(3)+2\pi\xi^{-1}\chi(q)
  +4\pi^2\xi^{-2}\,D\,\chi(q)\,,
  \eql{epz21}$$
where $q=e^{-\pi/\xi}$.

The low temperature form (\peq{epz21}) yields a somewhat simpler expression
for the free energy, (\peq{3free1}), as the $\ze_R(3)$ term goes out and
there is some cancellation that produces,
  $$
  \overline F_3={1\over240}+{\xi\over2\pi}D^2\,\chi(q)\,.
  $$
I have thus regained, at some length, the summation form (\peq{ssum})
with (\peq{sum9}).

All that has been done is to rederive, in a special case and in a
detailed way, the standard statistical mode sum, the general form of
which was obtained, in this fashion, in [\pref{DandK,Dow2}] and we are no
further forward in finding a closed form for the free energy, which is
one of my aims.

The equivalence of (\peq{epz2}) and (\peq{epz21}), derived here rather
trivially (granted Epstein's calculation), is a known inversion identity,
and can also be derived from (\peq{sm1}). The exact identity is written
out in Katayama [\pref{Kata}]. His derivation seems to be similar to that
via the Epstein function. Clearly the higher odd sphere expressions will
involve the Riemann \zf\ at positive odd integers. Katayama writes out
the one appropriate for the (partial) five-sphere.

Smart, [\pref{Smart}], also derives these identities, which is not
surprising since his work is concerned with the evaluation of the Epstein
\zf\ at integer arguments, through which he is led to the forms
$\phi_{2k}$. We see that Epstein had already arrived at the same
development. Consult also Bodendiek and Halbritter, [\pref{BandH}], for a
related treatment.

\section{\bf 7. The Epstein--Kober--Selberg--Chowla formula.}
Smart makes use of the, oft quoted, paper of Selberg and Chowla,
[\pref{SandC}], which is concerned, partly, with the evaluation of the
Epstein function via the Kronecker limit formula \footnote{Landau,
[\pref{Landau2}], gives some interesting history of this famous formula.}
and elegantly yields some of the finite forms for the complete elliptic
integral used in [\pref{DandK1}]. The important expression for the
Epstein function quoted by Selberg and Chowla, [\pref{SandC2}], and
derived by them in [\pref{SandC}], (equ.(6)), in terms of $K$--Bessel
functions, is essentially the same as that already given by Epstein,
[\pref{Epstein}], p.631, equn.(12), p.622, equn.(16). Rankin
[\pref{Rankin3}] obtained the same result and it also occurs in Bateman
and Grosswald, [\pref{BandG}]. These are the standard references. See
also Terras, [\pref{Terras}] p.209. Smart, [\pref{Smart}], employs this
formula to split the Epstein function into (simpler) holomorphic and
anti--holomorphic parts using the explicit form of the Bessel function,
$K_{t-1/2}$, which henceforth disappears from his analysis. Other
expressions were given by Deuring [\pref{Deuring}] and Mordell,
[\pref{Mordell}].

An early, and little known, derivation is that by Kober, [\pref{Kober}].
Following, and generalising, Watson, [\pref{Watson}], he obtains, {\it
starting} from the Bessel function end, the formula (I retain his
notation),
  $$\eqalign{
  {1\over\sqrt u}\,&\De^{(2w+1)/4}{\Ga(w+1/2)\over8\pi^{w+1/2}}Z_2(a,b,c;w+1/2)\cr
&={u^{-w}\over4}{\Ga(w)\over\pi^{w}}\,\ze_R(2w)
+{u^{w}\over4}{\Ga(w+1/2)\over\pi^{w+1/2}}\,\ze_R(2w+1)+\,\cr
  &\hspace{************}
  \sum_{n=1}^\infty \si_{2w}(n)n^{-w}\cos(2\pi vn)K_{w}(2\pi un)\,,}
  \eql{kob1}$$
where the quadratic form in the Epstein function, $Z_2$, is
$am_1^2+2bm_1m_2+cm_2^2$ and
  $$
  \De=ac-b^2\equiv a^2\,u^2\,,\quad v\equiv{b\over a}\,.
  $$
Equation (\peq{kob1}) is identical to the corresponding one in
[\pref{Epstein}] and [\pref{SandC}] and therefore, with justice, should
be called the Epstein--Kober formula. It is the {\it Fourier expansion}
of the Epstein function in $v$. For the sum of squares case, $v=0$, the
result is rederived by Guinand [\pref{Guinand2}] in the same way.

Kober also discusses behaviour under `reciprocal' transformations,
$u\to1/u$. In particular the diagonal case, $v=0$, when one can write
$u=\ssx$.

In this way of doing things, the Bessel function can be eliminated using
the standard formula
  $$
  \sum_{n=1}^\infty {\si_{2w}(n)\over n^{s+w}}=\ze_R(s+w)\,\ze_R(s-w)\,,
  \eql{sig}$$
and the Mellin transform (Heaviside's integral),
  $$
  {1\over4}\pi^{-s}\Ga\big({s+w\over2}\big)\,\Ga\big({s-w\over2}\big)=
  \int_0^\infty dy \,y^{s-1}\,K_w(2\pi y)\,,\quad \Real s>1+|\Real w|\,,
  \eql{hi}$$
so that,
  $$
  \sum_{n=1}^\infty {\si_{2w}(n)\over n^w} K_{w}(2\pi un)
=
{1\over2\pi i}\int_{c-i\infty}^{c+i\infty}ds\,u^{-s}\xi(s+w)\,\xi(s-w)\,,
  \eql{epmell}$$
with $c>1+w$ and where,
  $$\xi(2w)\equiv{1\over2}\pi^{-w}\,\Ga(w)\,\ze_R(2w)\,.
  $$
This is the same as Deuring's expression, [\pref{Deuring}]
p.589.

The integrand possesses a functional equation which follows most easily
from that for the Riemann \zf, $\xi(2w)=\xi(1-2w)$, and is,
  $$
  \xi(s+w)\,\xi(s-w)\equiv f_w(s)=f_w(1-s)\,.
  \eql{ffrel}$$

Using (\peq{ffrel}) one could {\it derive} the functional equation for the
Epstein \zf\ in (\peq{kob1}). The usual derivation employs the inversion
relation for generalised $\th$--functions, [\pref{Epstein,Krazer}].

Furthermore, it is possible to reverse the argument and use (\peq{epmell})
to obtain (\peq{kob1}) (for $v=0$). The diagonal Epstein function is
particularly easy to deal with being the \zf\ on the torus,
S$^1\times$S$^1$. It can be expressed very simply in terms of the \zfs\ on
the circle factors (these are Riemann \zfs\ ) and the result is exactly
(\peq{kob1}) with (\peq{epmell}) {\it without} the intervention of the
Bessel function. The individual Riemann \zfs\ in (\peq{kob1}) arise from
adjusting the zero modes. All this is standard and can easily be
generalised to higher dimensions.

A natural step is to proceed as with the holomorphic modular forms and ask
for the corresponding period polynomials which will follow directly from
(\peq{epmell}) and (\peq{ffrel}) in the standard manner.

There are four poles in the relevant strip, $-c\le s\le c$, and
a straightforward contour calculation produces the result,
  $$\eqalign{
  \sum_{n=1}^\infty {\si_{2w}(n)\over n^w}K_{w}(2\pi n u)-&
  {1\over u}\sum_{n=1}^\infty {\si_{2w}(n)\over n^w}K_{w}(2\pi n/u)\cr
  &={1\over2}\xi(2w)(u^{w-1}-u^{-w})+{1\over2}\xi(-2w)(u^{-w-1}-u^{w})\,,}
  \eql{ppk}$$
The right hand side of (\peq{ppk}) might be termed a period function.

This formula was obtained by Guinand, [\pref{Guinand2}], but using the
complete equation (\peq{kob1}), with $v=0$, and properties of the Epstein
function. This is unnecessary as I have just shown.

In order to make a connection with the period polynomials encountered
earlier,\eg (\peq{are1}), one introduces the formula for the $K$-Bessel
function, written as,
  $$
  K_{t-1/2}(x)=\sqrt{2\over \pi}\,x^{t-1/2}\,(-1)^{t-1}\,\bigg({1\over x}
{d\over dx}\bigg)^{t-1}{e^{-x}\over x}\,,\quad t\in\oZ\,.
  \eql{kbess}$$
Then, somewhat similarly to [\pref{Smart}],
  $$\eqalign{
  \sum_{n=1}^\infty {\si_{2t-1}(n)\over n^{t-1/2}}K_{t-1/2}(2\pi n u)&=
  \sqrt{2\over\pi}{(-1)^{t-1}\over(2\pi)^{t-1/2}}\bigg({1\over u}
  {d\over du}\bigg)^{t-1}
  \sum_{n=1}^\infty{\si_{2t-1}(n)\over n^{2t-1}}\,q^{2n}\cr
  &=\sqrt{2\over\pi}{(-1)^{t-1}\over(2\pi)^{t-1/2}}
  \bigg({1\over u}{d\over du}\bigg)^{t-1}
  S_t(iu)\,,}
  $$
so that, for the left--hand side of (\peq{ppk}),  one finds,
  $$
  (-1)^{t-1}{1\over{\pi}^t}\bigg(\bigg({d\over du^2}\bigg)^{t-1}
  S_t(iu)-{1\over u}\bigg({d\over d(1/u^2)}\bigg)^{t-1}S_t(i/u)\bigg)\,.
 \eql{krelp} $$
It is not obvious from (\peq{are1}) that this reduces so nicely to
(\peq{ppk}), and I leave this as a question mark.

Further expressions involving Bessel functions appear in the Appendix.
\begin{ignore}

This can be turned around in the following way. Write (\peq{kbess})
explicitly in the form,\mgn{THis is wrong}
  $$
 e^{-x}\sum_{j=0}^{t-1}
  {(t-1+j)!\over j!(t-1-j)!}\,(2x)^{-j-1/2}=\pi^{-1/2}\,K_{t-1/2}(x)\,,
  $$
and solve this for $(2x)^{-j-1/2}$, $0\le j\le t-1$,
  $$
  e^{-x}\,(2x)^{-j-1/2}={1\over\sqrt\pi}\sum_{t=0}^{j-1}{()!\over()!}
  \,K_{t-1/2}(x)\,.
  $$
Now set $x=2\pi n u$ and get
  $$
  {e^{-2\pi nu}\over n^{j+1/2}}=(2\pi)^{j+1/2}
  {1\over\sqrt\pi}\sum_{t=0}^{j-1}{()!\over()!}\,K_{t-1/2}(2\pi nu)
  $$
\end{ignore}
\begin{ignore}
\section{\bf More hyperbolic sums.}
In terms of the hyperbolic expression, (\peq{speen}), the quantity
corresponding to the $c_t$ is,
  $$
  {\pa\over\pa\beta}\,E(\beta)=-{1\over2^{d+1}}\sum_{m=1}^\infty m\bigg({(d-1)^2
\cosh(m\be/2)\over\sinh^{d-1}(m\be/2)}+{d(d+1)\cosh(m\be/2)
\over\sinh^{d+1}(m\be/2)}\bigg)\,,
  \eql{speen2}$$
which means that it is also possible to evaluate the sums,
  $$
  \sum_{m=1}^\infty {m^{2p+1}\cosh(mx)\over\sinh^d(mx)}
  $$
and
  $$
  \sum_{m=1}^\infty {m^{2p}\over\sinh^d(mx)}\,,
  $$
at special values of $x$. Cases with $d=1$ have been considered by Zucker,
[\pref{zucker2}] and Ling, [\pref{Ling2}] and can follow from expansion of
the dn function.
\section{\bf The fermion case. More to be done.}

In [\pref{DandK1}] it was shown that just as for bosons, the fermion
internal energies on all spheres can be expressed in terms of just two
quantities which could be taken to be the first two partial energies,
$\eta_1$ and $\eta_2$. This was proved without recourse to the boson result
and in a different manner. The recursion is not similar to (\peq{recurs})
but has a cubic form. One might therefore reasonably expect an equation
corresponding to (\peq{shr}) to exist, but in a different form and derived
by a different route, possibly from the Jacobi function ds $u$.

A preliminary fact, that it is useful to confirm, starts from the
expression for the partial spinor energy on the circle given in
[\pref{DandK1}],
  $$
  \eta_1(\be)=-{1\over24} \gK^2\,(1-2k^2)\,,
  $$
which, using (\peq{diffrel}), can be integrated to give,
  $$
   \int d\be\, \eta_1(\be)={1\over12}\int d\ka {1-2\ka\over\ka(1-\ka)}=
  {1\over12}\log(\ka\ka'/16)\,,
  $$
where $\ka=k^2$ and a constant of integration has been chosen to make the
free energy and internal energies agree at absolute zero.

I have, in this roundabout way, derived another of Jacobi's formulae,
  $$
  q^{1/24}\prod_{j=0}^\infty(1+q^{2j+1})=2^{1/6}(kk')^{-1/12}\,,
  $$
the logarithm of which gives the mode sum form of the circle fermion
free energy, as is well known in various contexts.

We can also see the difficulty if we attempt the same approach for $\eta_2$
  $$\eqalign{
   -\int d\be\,\eta_2(\be)&={1\over120}\int d\ka\,\gK^2(\ka)\,
{(7+8\ka-8\ka^2)\over\ka(1-\ka)}\cr
&={1\over120}\int d\ka\,\gK^2(\ka)\,\bigg({7\over\ka}
+{7\over\ka'}-8\bigg)\,,}
  $$
and the integral is not obvious.  I merely note the hypergeometric
representation,
  $$
  \gK(\ka)=F\big({1\over2},{1\over2},1; \ka\big)={1\over\pi\ka^{1/4}}\,
  Q_{-1/2}\big((\ka+1)/2\sqrt\ka\big)\,.
  $$
A similar thing holds in the boson case of course.
\end{ignore}
\section{\bf 8. Conclusion}

It does not seem possible to ellipticise integrals of the internal
energy, in particular the free energy (except for the circle), because
the modular properties are more complicated possessing nonzero cocycle
functions. These are related to a known Lambert series connected with the
Eisenstein series and Zagier's expression for the period functions of
these is more neatly re--computed by a contour method. The use of the
fully subtracted series $\ep_t^{\rm sub}$ allows a more compact
treatment.

Sufficient historical material has been exhibited to indicate that the
Selberg--Chowla formula should be renamed the Epstein--Kober formula.
\newpage
\section{\bf Appendix, Dirichlet Series.}

It is possible to give a fairly broad context to the inversion behaviour
(\eg modular properties), without too much restriction, in terms of
Dirichlet series. Although these ideas, by now classic, are associated
with Hecke, and have been expounded by Ogg [\pref{Ogg,Ogg2}], for
example, the simplest case occurs earlier in other places. I mention
Koshliakov, [\pref{Kosh}], whose discussion I now paraphrase both for
historical justice and interest.

Define two ordinary Dirichlet series (`zeta functions'),
  $$
  \phi(s)=\sum_{n=1}^\infty {a_n\over n^s}\,,\,\,(\Real s>\nu_a)\,,\quad
  \psi(s)=\sum_{n=1}^\infty {b_n\over n^s}\,,\,\,(\Real s>\nu_b>\nu_a)\,,
  $$
for some $\nu$'s depending on the asymptotics of $a_n$ and $b_n$, and
{\it assume} that the $a_n$ and $b_n$ are such that the corresponding
`cylinder kernels' satisfy the `formula of transformation',
  $$
  \sum_{n=0}^\infty a_n\,e^{-nb\rho}={a\over\rho^\nu}
  \sum_{n=0}^\infty b_n\,e^{-nb/\rho}\,,\quad \nu=\nu_a\,,\quad \rho>0\,,
  \eql{trans2}$$
for fixed $a$ and $b$, where the numbers (not necessarily integers)
of `zero modes' are taken to be,
  $$
  a_0=-\phi(0)\,\quad b_0=-\psi(0)\,.
  $$
Then Koshliakov, [\pref{Kosh}], shows the equivalence of (\peq{trans2})
with
  $$
  a\,{\Ga(\nu-s)\,\psi(\nu-s)\over b^{\nu-s}}={\Ga(s)\,\phi(s)\over b^s}\,,
  \eql{frel}$$
by brute force using the summation function, $\si(z)$, given by the
Bessel function expression,
  $$
  \si(z)=-2ab\,z^{(\nu-1)/2}\,\sum_{n=1}^\infty {b_n\over n^{(\nu-1)/2}}\,
  K_{\nu-1}\big(2b\sqrt{nz}\big)\,,
  $$
and Heaviside's formula, (\peq{hi}). $\si(z)$ has the important property
of possessing poles at $z=1,2,\ldots$ with residues $a_1,a_2,\ldots$ and
also has a cut along the negative $x$--axis with discontinuity
$ab^\nu2\pi i\psi(0)(-x)^{\nu-1}/\Ga(\nu)$, $z=x+iy$.

Koshliakov shows that $\phi(s)$ is a single--valued holomorphic function
having a first-order pole at $s=\nu$,
  $$
  \phi(s)\sim-\psi(0){ab^\nu\over\Ga(\nu)}{1\over s-\nu}\,,
  \eql{phipole}$$
and it is then a theorem that (\peq{frel}) together with (\peq{phipole}) is
equivalent to (\peq{trans2}), as can also be established easily, and
instructively, by Mellin transform. It follows, symmetrically, that
$\psi(s)$ is a single--valued holomorphic function having a first-order
pole at $s=\nu$,
  $$
  \psi(s)\sim-\phi(0){b^\nu\over a\Ga(\nu)}{1\over s-\nu}\,.
  $$

Koshliakov also gives an integral form of $\phi(s)$,
  $$
  \phi(s)=-{\sin\pi s\over\pi}\int_0^\infty dx\, {\si(x)\over x^s}
  \,,\quad \Real s<0\,,
  \eql{intphi}$$
showing that $\phi(s)$ vanishes at negative integers in agreement with
(\peq{frel}).

Hecke, [\pref{Hecke}], derives the same equivalence, but only for
$a_n=b_n$, and eight years later. The more general case can also be found
in the standard references Ogg [\pref{Ogg,Ogg2}] and Weil,
[\pref{Weil2}], and no doubt elsewhere. A related formula involving
Ramanujan's `reciprocal functions' occurs in Hardy and Littlewood,
[\pref{HandL}].

Defining cylinder--kernels, or theta--series, including the zero modes,
by,
  $$
  \Phi(\be)=\sum_{n=0}^\infty a_n\,e^{-n\be}\,,\quad
  \Psi(\be)=\sum_{n=0}^\infty b_n\,e^{-n\be}\,,
  $$
the reciprocal relation, (\peq{trans2}), reads,
  $$
  \Phi(b\rho)={a\over\rho^\nu}\,\Psi(b/\rho)\,,
  \eql{recip}
  $$
(continued from the regions of convergence in $s$) and one has, in
standard fashion (\cf (\peq{ell})),
  $$
  \phi(s)={1\over\Ga(s)}\int_0^\infty d\be\,\be^{s-1}\,\big(\Phi(\be)-a_0\big)\,,
  \quad
  \psi(s)={1\over\Ga(s)}\int_0^\infty d\be\,\be^{s-1}
  \,\big(\Psi(\be)-b_0\big)\,.
  $$

As an example, the Dirichlet series corresponding to the classical
Eisenstein series is, from (\peq{mell1}) or (\peq{sig}),
  $$
  \sum_{m=1}^\infty \si_{2t-1}(m)\,m^{-s}=\ze_R(s)\,\ze_R(s+1-2t)\,,
  $$
and this is a case discussed by Koshliakov, [\pref{Kosh}] p.19, with
$a_n=b_n=\si_{2t-1}(n), \nu=2t,a=(-1)^t,b=2\pi,\phi(0)=\psi(0)=-B_{2t}/4t$.

Koshliakov also considers Jacobi elliptic cases that yield reciprocal
identities covered earlier by Glaisher [\pref{Glaisher}].

\subsection{\it Generalised Dirichlet series.}
I have described above the classic Dirichlet set up. A certain, and
sometimes only apparent, generalisation is obtained by replacing $n$ by
arbitrary sequences, $\la_m$ and $\mu_n$, and defining new $\phi$ and
$\psi$ by
  $$
  \phi(s)=\sum_{m=1}^\infty {a_m\over \la_m^s}\,,\quad\quad
  \psi(s)=\sum_{n=1}^\infty {b_n\over \mu_n^s}\,,
  \eql{newdir}$$
which satisfy, {\it by definition}, the functional equation
(one of many possible),
  $$
  \Ga(\de-s)\,\psi(\de-s)=\Ga(s)\,\phi(s)\,,\quad \de\in\oR\,,
  \eql{frel2}$$
with corresponding heat--kernels,
  $$
  \Phi(\be)=\sum_{m=1}^\infty a_m\,e^{-\la_m\be}\,,\quad
  \Psi(\be)=\sum_{n=1}^\infty b_n\,e^{-\mu_n\be}\,.
  \eql{hkz}$$

Bochner, [\pref{Bochner}], (see also Chandrasekharan and Narasimhan,
[\pref{CandN}], Knopp, [\pref{Knopp2}] ), derives, by Mellin transforms,
the `modular relation',
  $$
  \Phi(\be)-\be^{-\de}\,\Psi\big({1\over\be}\big)=B(\be)\,,
  \eql{modrel}$$
where the `residual function' $B(\be)$ is
  $$
  B(\be)={1\over2\pi i}\int_{C(\caS)} ds\,\chi(s)\,\be^{-s}=\sum_{s\in\caS}\be^{-s}
  {\rm Res}\,\chi(s)\,.
  $$
$\chi(s)$ is the joint continuation of the two sides of (\peq{frel2}) and
is assumed to have only simple poles for singularities confined to a
compact region, $\caS$, of the $s$--plane, usually on the real axis.
$C(\caS)$ is a loop surrounding $\caS$. I especially note that if there
exist higher--order poles then integer powers of $\log \be$ occur.

In certain situations, \eg Koshliakov's, $B(\be)$ can be naturally
amalgamated with
the left--hand side to give an exact modular relation like (\peq{trans2}),
\eg Bochner,
[\pref{Bochner}] p.342. Typically one would include zero modes $\la_0=0$
and $\mu_0=0$ in (\peq{hkz}), but not in (\peq{newdir}). (There seems to be
a misprint on p.342 of [\pref{Bochner}] which has $\la_0=\mu_0=1$.)

There is another formula connected with this analysis that is useful. It
is concerned with the representation of the {\it modified} series,
  $$
  \phi(s,w)=\sum_{n=1}^\infty {a_n\over (\la_n+w^2)^s}\,\quad\Real w>0.
  \eql{asser}$$

There are many reasons why we should wish to add the number $w^2$. For
example it might correspond to a mass or simply be added for extra
flexibility as when calculating heat--kernel expansion coefficients via
resolvents or it might be another part of the denominator as in Epstein's
derivation of (\peq{kob1}); see [\pref{DandKi}].

The representation alluded to is, (Berndt [\pref{Berndt4}]),
  $$
  \Ga(s)\,\phi(s,w)=R(s,w)+2\sum_{n=1}^\infty b_n\,\bigg({\mu_n\over w^2}
  \bigg)^{(s-\de)/2}\,K_{s-\de}\big(2w\sqrt\mu_n\big)\,,
  \eql{berndt}$$
where $s$ is such that the summation converges absolutely and $R(s,w)$ is
given by
  $$
  R(s,w)=\sum_{s'\in\,\caS} \Ga(s-s')\,w^{2s'-2s}{\rm Res}\,\, \chi(s')\,.
  \eql{resi}$$

Berndt's second proof of (\peq{berndt}) proceeds via the
cylinder--kernels, $\Phi$ and $\Psi$. I prefer his first, Mellin, proof,
[\pref{Berndt6}]. Of course, the information employed is the same. I give
this proof out of interest.

The Mellin transform relation between $\phi(s)$ and the $\phi(s,w)$ of
(\peq{asser}), is
  $$
  \Ga(s)\,\phi(s,w)={1\over2\pi i}\int_{c-i\infty}^{c+i\infty}ds'\,w^{2s'-2s}
  \Ga(s-s')\Ga(s')\,\phi(s')
  $$
where $c$ is chosen sufficiently positive to make both $\phi(s)$ and
$\psi(s)$ converge. The vertical contour is now moved to the left so as
to run from $\de-c-i\infty$ to $\de-c+i\infty$ picking up the residues,
$R(s,w)$, (\peq{resi}), on the way. The region, $\caS$, lies between
these vertical lines and also the horizontal pieces contribute zero.

On the shifted vertical contour make the coordinate change $s'\to\de-s'$
and use (\peq{frel2}) \ie $\chi(\de-s')=\Ga(s')\psi(s')$, valid in this
range, and so for this part I find,
  $$\eqalign{
  \sum_n b_n {1\over2\pi i}&\int_{c-i\infty}^{c+i\infty}ds'\,\Ga(s'-\de+s)
  \Ga(s')\mu_n^{-s'} w^{2\de-2s'-2s}\cr
&=2\sum_{n=1}^\infty b_n\,\bigg({\mu_n\over w^2}
  \bigg)^{(s-\de)/2}\,K_{s-\de}\big(2w\sqrt\mu_n\big)\,,}
  $$
on using (\peq{hi}). Hence (\peq{berndt}) has been established.

In specific situations (\peq{berndt}) is quite familiar in the \zf\ area
and in  background and finite--temperature quantum field theory.

An illustrative example is the simplest, diagonal Epstein function,
   $$
   \phi(s)=Z_p(s)\equiv\sumdasht{{\bf m}=-\infty}{\infty}
  {1\over({\bf m.m})^s}\,,\quad {\bf m}=(m_1,m_2,\ldots,m_p)\,,
   \eql{diagep}$$
which obeys the functional equation (Epstein [\pref{Epstein}]),
   $$
   \Ga(s)\,Z_p(s)=\pi^{2s-p/2}\,\Ga(p/2-s)\,Z_p(p/2-s)\,,
   \eql{epfunc}$$
so that, see (\peq{frel2}), $\de=p/2$, $a_n=1$, $\la_m = {\bf m.m}$,
$b_n=\pi^{p/2}$, $\mu_n=\pi^2\,{\bf m.m}$.

$\Ga(s)\,Z_p(s)$ has poles at $s=0$ and $s=p/2$ and (\peq{berndt}) gives
for the modified, or inhomogeneous, or `massive' Epstein function,
$Z_p(s,w)$,
  $$\eqalign{
  \sumdasht{{\bf m}=-\infty}{\infty}
  {1\over({\bf m.m}+w^2)^s}=-w^{-2s}&+{\Ga(p/2)\Ga(s-p/2)\over\Ga(s)}
  \,w^{p-2s}\cr
 & +{2\pi^s\over\Ga(s)}\sumdasht{{\bf m}=-\infty}{\infty}
  \bigg({|{\bf m}|\over w}\bigg)^{s-p/2}\,
  K_{s-p/2}\big(2w\pi|{\bf m}|\big)\,,}
  \eql{modep}$$
where the dash means to omit the ${\bf m=0}$ term. Note that the $w^{-2s}$ could be
included in the left--hand sum if this were extended to include ${\bf m=0}$.

The right--hand side can be taken as the continuation of the left--hand
side showing the single pole at $s=p/2$, as is correct and may be proved in
other ways.

Actually (\peq{modep}) is shown much more directly using the Jacobi
inversion relation, which is, of course, how Epstein obtained
(\peq{epfunc}). Equivalent to Jacobi inversion is Poisson summation.

Equations like (\peq{modep}) also occur in lattice summations. Some references,
but none earlier than those already quoted, are given in the review by Glasser
and Zucker [\pref{GandZ}] p.109.

A similar formula holds for the general Epstein function which may, or may
not, have a pole at $s=p/2$ and may, or may not, vanish at $s=0$. The works
[\pref{Kirsten1,Elizalde2,Elizalde3}], for example, can be consulted for details,
some further references and physical applications.

When $p=1$, $\phi(s)=2\ze_R(2s)$ and I recover an earlier, well--known
formula, [\pref{Watson,Kober}], which yields, after putting $w=u\,m_2$
and summing over $m_2$, the formula for the Epstein \zf\ mentioned
earlier; see (\peq{kob1}).

It is left as an exercise to show that the non-diagonal formula,
(\peq{kob1}), can be obtained in the same way using  reciprocal relations
for the Hurwitz \zf, or, equivalently, the Lipshitz formula (\cf Epstein,
[\pref{Epstein}], Kober [\pref{Kober}] p.622).

As an application of (\peq{modep}) it might be helpful if I present a
standard derivation of the usual statistical free energy mode sum from
the thermal \zf\ expression, for a reasonably general system. The
starting point is the determinant form,
  $$
  F={i\over2}\lim_{s\to0}{\ze(s,\be)\over s}
  =-{1\over2\be}\lim_{s\to0}{1\over s}
  \sumdasht{{m=-\infty\atop n}}
{\infty}{d_n\over
 \big(\om_n^2+4\pi^2m^2/\be^2\big)^s}\,,
  \eql{eff2}$$
where $\om_n^2$ and $d_n$ are the eigenvalues and degeneracies of the
appropriate operator (related to the Laplacian) on the spatial section,
$\man$. The $\om_n$ are the single--particle energies. The dash here means
that the denominator should never be zero (\cf (\peq{eff})) but the sum
includes $m=0$. For simplicity it is assumed that there is no spatial zero
mode and so the dash can be removed.

Now consider (\peq{modep}) for $p=1$. Include the $w^{-2s}$ with the
left--hand sum, set $w=\be\om_n/2\pi$, multiply by the degeneracy, $d_n$
and sum over $n$. The limit (\peq{eff2}) is then easily taken since
$s\Gamma(s)=\Gamma(s+1)$ and only the $K_{1/2}$ Bessel appears. There is
nothing deep about this calculation and the result is a standard form,
  $$\eqalign{
  F=&{1\over2}\,\ze_\man(-1/2)+{1\over\be}\sum_n d_n
  \sum_{m=1}^\infty{e^{-m\om_n\be}\over m}\cr
  =&{1\over2}\,\ze_\man(-1/2)+{1\over\be}\sum_n d_n\log(1-e^{-\om_n\be})\,.}
  \eql{modesum}$$
To avoid specifically field theoretic problems, it has been assumed that
$\ze_\man(-1/2)$ exists. This is true for the cases discussed in this
paper.

The result, (\peq{modesum}), is an analogue of the (first) Kronecker
limit--formula applied to the generalised torus S$^1\times\man$; see
[\pref{DandA}]. The original formula relates to $\man=$S$^1$ and, in this
case, there exists a functional relation, (\peq{frel2}), so that the
limit--formula is often expressed in terms of the remainder about the
pole in $\ze(s,\be)$ at $s=1$. In the general case this is not possible.

Finally, if one introduces the standard arithmetic quantity $r_p(n)$,
equal to the number of representations of the integer $n$ as the sum of
$p$ squares, then the diagonal Epstein function, (\peq{diagep}), is
obtained by writing,
  $$
  \phi(s)=Z_p(s)=\sum_n{r_{p}(n)\over n^s}\,,
  $$
and is treated as such by Koshliakov who also gives the `exact', product
forms of $Z_p(s)$ for $p=2$ and $p=4$.

\begin{ignore}
The other useful formula is a generalisation of one occurring in the
theory of Riesz means which has already been encountered, essentially in
(\peq{mint}) and (\peq{soln}).

With the same notation as above, one has, for $h\ge0$,
  $$
  {1\over\Ga(h+1)}\sum_{\la_n}a_n\,\th(x-\la_n)\,(x-\la_n)^h
  ={1\over2\pi i}\int_{c-i\infty}
  ^{c+i\infty}ds\,{\Ga(s)\,\phi(s)\,x^{s+h}\over\Ga(h+1+s)}\,,
  $$
where $\th$ is Heaviside's step function.
\end{ignore}
\newpage
\section{\bf References.}

\begin{putreferences}
   \ref{RandA}{Rao,M.B. and Ayyar,M.V. \jims{15}{1923/24}{150}.}
  \ref{DandK2}{Dowker,J.S. and Kirsten,K. {\it Elliptic aspects of
  statistical mechanics on spheres} ArXiv:0807.1995.}
  \ref{DandK1}{Dowker,J.S. and Kirsten,K. \np {638}{2002}{405}.}
  \ref{GandS}{Gel'fand, I.M. and Shilov, G.E. {\it Generalised Functions}, vol.1
  (Academic Press, New York, 1964). }
  \ref{DandA}{Dowker,J.S. and Apps, J.S. \cqg{12}{1995}{1363}.}
  \ref{Weil}{Weil,A., {\it Elliptic functions according to Eisenstein and
  Kronecker}, Springer, Berlin, 1976.}
  \ref{Ling}{Ling,C-H. {\it SIAM J.Math.Anal.} {\bf5} (1974) 551.}
  \ref{Ling2}{Ling,C-H. {\it J.Math.Anal.Appl.}(1988).}
 \ref{BMO}{Brevik,I., Milton,K.A. and Odintsov, S.D. {\it Entropy bounds in
 $R\times S^3$ geometries}. hep-th/0202048.}
 \ref{KandL}{Kutasov,D. and Larsen,F. {\it JHEP} 0101 (2001) 1.}
 \ref{KPS}{Klemm,D., Petkou,A.C. and Siopsis {\it Entropy
 bounds, monoticity properties and scaling in CFT's}. hep-th/0101076.}
 \ref{DandC}{Dowker,J.S. and Critchley,R. \prD{15}{1976}{1484}.}
 \ref{AandD}{Al'taie, M.B. and Dowker, J.S. \prD{18}{1978}{3557}.}
 \ref{Dow1}{Dowker,J.S. \prD{37}{1988}{558}.}
 \ref{Dow3}{Dowker,J.S.\prD{28}{1983}{3013}.}
 \ref{DandK}{Dowker,J.S. and Kennedy,G. \jpa{11}{1978}{895}.}
 \ref{Dow2}{Dowker,J.S. \cqg{1}{1984}{359}.}
 \ref{DandKi}{Dowker,J.S. and Kirsten, K. {\it Comm. in Anal. and Geom.
  }{\bf7}(1999) 641.}
 \ref{DandKe}{Dowker,J.S. and Kennedy,G.
 \jpa{11}{1978}{895}.} \ref{Gibbons}{Gibbons,G.W. \pl{60A}{1977}{385}.}
 \ref{Cardy}{Cardy,J.L. \np{366}{1991}{403}.}
 \ref{ChandD}{Chang,P. and
  Dowker,J.S. \np{395}{1993}{407}.}
  \ref{DandC2}{Dowker,J.S. and
  Critchley,R. \prD{13}{1976}{224}.}
 \ref{Camporesi}{Camporesi,R.
 \prp{196}{1990}{1}.}
 \ref{BandM}{Brown,L.S. and Maclay,G.J.
 \pr{184}{1969}{1272}.}
 \ref{CandD}{Candelas,P. and Dowker,J.S.
 \prD{19}{1979}{2902}.}
  \ref{Unwin1}{Unwin,S.D. Thesis. University of
 Manchester. 1979.} \ref{Unwin2}{Unwin,S.D. \jpa{13}{1980}{313}.}
 \ref{DandB}{Dowker,J.S. and Banach,R. \jpa{11}{1979}{}.}
 \ref{Obhukov}{Obhukov,Yu.N. \pl{109B}{1982}{195}.}
 \ref{Kennedy}{Kennedy,G. \prD{23}{1981}{2884}.}
 \ref{CandT}{Copeland,E.
 and Toms,D.J. \np {255}{1985}{201}.} \ref{ELV}{Elizalde,E., Lygren, M.
 and Vassilevich, D.V. \jmp {37}{1996}{3105}.}
 \ref{Malurkar}{Malurkar,S.L. {\it J.Ind.Math.Soc} {\bf16} (1925/26) 130.}
 \ref{Glaisher}{Glaisher,J.W.L. {\it Messenger of Math.} {\bf18} (1889)
  1.}
 \ref{Anderson}{Anderson,A. \prD{37}{1988}{536}.}
 \ref{CandA}{Cappelli,A. and D'Appollonio,\pl{487B}{2000}{87}.}
 \ref{Wot}{Wotzasek,C. \jpa{23}{1990}{1627}.}
 \ref{RandT}{Ravndal,F. and Tollesen,D. \prD{40}{1989}{4191}.}
 \ref{SandT}{Santos,F.C. and Tort,A.C. \pl{482B}{2000}{323}.}
 \ref{FandO}{Fukushima,K. and Ohta,K. {\it Physica} {\bf A299} (2001) 455.}
 \ref{GandP}{Gibbons,G.W. and Perry,M. \prs{358}{1978}{467}.}
 \ref{Dow4}{Dowker,J.S. {\it Zero modes, entropy bounds and partition
 functions.} hep-th\break /0203026.}
  \ref{Rad}{Rademacher,H. {\it Topics in analytic number theory,}
 (Springer- Verlag,  Berlin, 1973).}
  \ref{Halphen}{Halphen,G.-H. {\it Trait\'e des Fonctions Elliptiques}, Vol 1,
  Gauthier-Villars, Paris, 1886.}
  \ref{CandW}{Cahn,R.S. and Wolf,J.A. {\it Comm.Mat.Helv.} {\bf 51} (1976) 1.}
  \ref{Berndt}{Berndt,B.C. \rmjm{7}{1977}{147}.}
  \ref{Hurwitz}{Hurwitz,A. \ma{18}{1881}{528}.}
  \ref{Hurwitz2}{Hurwitz,A. {\it Mathematische Werke} Vol.I. Basel,
  Birkhauser, 1932.}
  \ref{Berndt2}{Berndt,B.C. \jram{303/304}{1978}{332}.}
  \ref{RandA}{Rao,M.B. and Ayyar,M.V. \jims{15}{1923/24}{150}.}
  \ref{Hardy}{Hardy,G.H. \jlms{3}{1928}{238}.}
  \ref{TandM}{Tannery,J. and Molk,J. {\it Fonctions Elliptiques},
   Gauthier-Villars, Paris, 1893--1902.}
  \ref{schwarz}{Schwarz,H.-A. {\it Formeln und Lehrs\"atzen zum Gebrauche..},
  Springer 1893.(The first edition was 1885.) The French translation by
  Henri Pad\'e is, Gauthier-Villars, Paris, 1894}
  \ref{Hancock}{Hancock,H. {\it Theory of elliptic functions}, Vol I.
   Wiley, New York 1910.}
  \ref{watson}{Watson,G.N. \jlms{3}{1928}{216}.}
  \ref{MandO}{Magnus,W. and Oberhettinger,F. {\it Formeln und S\"atze},
  Springer-Verlag, Berlin 1948.}
  \ref{Klein}{Klein,F. {\it }.}
  \ref{AandL}{Appell,P. and Lacour,E. {\it Fonctions Elliptiques},
  Gauthier-Villars,
  Paris, 1897.}
  \ref{HandC}{Hurwitz,A. and Courant,C. {\it Allgemeine Funktionentheorie},
  Springer,
  Berlin, 1922.}
  \ref{WandW}{Whittaker,E.T. and Watson,G.N. {\it Modern analysis},
  Cambridge 1927.}
  \ref{SandC}{Selberg,A. and Chowla,S. \jram{227}{1967}{86}. }
  \ref{zucker}{Zucker,I.J. {\it Math.Proc.Camb.Phil.Soc} {\bf 82 }(1977) 111.}
  \ref{glasser}{Glasser,M.L. {\it Maths.of Comp.} {\bf 25} (1971) 533.}
  \ref{GandW}{Glasser, M.L. and Wood,V.E. {\it Maths of Comp.} {\bf 25} (1971)
  535.}
  \ref{greenhill}{Greenhill,A,G. {\it The Applications of Elliptic
  Functions}, MacMillan, London, 1892.}
  \ref{Weierstrass}{Weierstrass,K. {\it J.f.Mathematik (Crelle)}
  {\bf 52} (1856) 346.}
  \ref{Weierstrass2}{Weierstrass,K. {\it Mathematische Werke} Vol.I,p.1,
  Mayer u. M\"uller, Berlin, 1894.}
  \ref{Fricke}{Fricke,R. {\it Die Elliptische Funktionen und Ihre Anwendungen},
    Teubner, Leipzig. 1915, 1922.}
  \ref{Konig}{K\"onigsberger,L. {\it Vorlesungen \"uber die Theorie der
 Elliptischen Funktionen},  \break Teubner, Leipzig, 1874.}
  \ref{Milne}{Milne,S.C. {\it The Ramanujan Journal} {\bf 6} (2002) 7-149.}
  \ref{Schlomilch}{Schl\"omilch,O. {\it Ber. Verh. K. Sachs. Gesell. Wiss.
  Leipzig}  {\bf 29} (1877) 101-105; {\it Compendium der h\"oheren Analysis},
  Bd.II, 3rd Edn, Vieweg, Brunswick, 1878.}
  \ref{BandB}{Briot,C. and Bouquet,C. {\it Th\`eorie des Fonctions Elliptiques},
  Gauthier-Villars, Paris, 1875.}
  \ref{Dumont}{Dumont,D. \aim {41}{1981}{1}.}
  \ref{Andre}{Andr\'e,D. {\it Ann.\'Ecole Normale Superior} {\bf 6} (1877) 265;
  {\it J.Math.Pures et Appl.} {\bf 5} (1878) 31.}
  \ref{Raman}{Ramanujan,S. {\it Trans.Camb.Phil.Soc.} {\bf 22} (1916) 159;
 {\it Collected Papers}, Cambridge, 1927}
  \ref{Weber}{Weber,H.M. {\it Lehrbuch der Algebra} Bd.III, Vieweg,
  Brunswick 190  3.}
  \ref{Weber2}{Weber,H.M. {\it Elliptische Funktionen und algebraische Zahlen},
  Vieweg, Brunswick 1891.}
  \ref{ZandR}{Zucker,I.J. and Robertson,M.M.
  {\it Math.Proc.Camb.Phil.Soc} {\bf 95 }(1984) 5.}
  \ref{JandZ1}{Joyce,G.S. and Zucker,I.J.
  {\it Math.Proc.Camb.Phil.Soc} {\bf 109 }(1991) 257.}
  \ref{JandZ2}{Zucker,I.J. and Joyce.G.S.
  {\it Math.Proc.Camb.Phil.Soc} {\bf 131 }(2001) 309.}
  \ref{zucker2}{Zucker,I.J. {\it SIAM J.Math.Anal.} {\bf 10} (1979) 192,}
  \ref{BandZ}{Borwein,J.M. and Zucker,I.J. {\it IMA J.Math.Anal.} {\bf 12}
  (1992) 519.}
  \ref{Cox}{Cox,D.A. {\it Primes of the form $x^2+n\,y^2$}, Wiley, New York,
  1989.}
  \ref{BandCh}{Berndt,B.C. and Chan,H.H. {\it Mathematika} {\bf42} (1995) 278.}
  \ref{EandT}{Elizalde,R. and Tort.hep-th/}
  \ref{KandS}{Kiyek,K. and Schmidt,H. {\it Arch.Math.} {\bf 18} (1967) 438.}
  \ref{Oshima}{Oshima,K. \prD{46}{1992}{4765}.}
  \ref{greenhill2}{Greenhill,A.G. \plms{19} {1888} {301}.}
  \ref{Russell}{Russell,R. \plms{19} {1888} {91}.}
  \ref{BandB}{Borwein,J.M. and Borwein,P.B. {\it Pi and the AGM}, Wiley,
  New York, 1998.}
  \ref{Resnikoff}{Resnikoff,H.L. \tams{124}{1966}{334}.}
  \ref{vandp}{Van der Pol, B. {\it Indag.Math.} {\bf18} (1951) 261,272.}
  \ref{Rankin}{Rankin,R.A. {\it Modular forms} (CUP, Cambridge, 1977).}
  \ref{Rankin2}{Rankin,R.A. {\it Proc. Roy.Soc. Edin.} {\bf76 A} (1976) 107.}
  \ref{Skoruppa}{Skoruppa,N-P. {\it J.of Number Th.} {\bf43} (1993) 68 .}
  \ref{Down}{Dowker.J.S. \np {104}{2002}{153}.}
  \ref{Eichler}{Eichler,M. \mz {67}{1957}{267}.}
  \ref{Zagier}{Zagier,D. \invm{104}{1991}{449}.}
   \ref{KandZ}{Kohnen, W. and Zagier,D. {\it Modular forms with rational periods.}
   in {\it Modular Forms} ed. by Rankin, R.A. (Ellis Horwood, Chichester, 1984).}
  \ref{Lang}{Lang,S. {\it Modular Forms}, (Springer, Berlin, 1976).}
  \ref{Kosh}{Koshliakov,N.S. {\it Mess.of Math.} {\bf 58} (1928) 1.}
  \ref{BandH}{Bodendiek, R. and Halbritter,U. \amsh{38}{1972}{147}.}
  \ref{Smart}{Smart,L.R., \pgma{14}{1973}{1}.}
  \ref{Grosswald}{Grosswald,E. {\it Acta. Arith.} {\bf 21} (1972) 25.}
  \ref{Kata}{Katayama,K. {\it Acta Arith.} {\bf 22} (1973) 149.}
  \ref{Ogg}{Ogg,A. {\it Modular forms and Dirichlet series} (Benjamin, New York,
   1969).}
  \ref{Bol}{Bol,G. \amsh{16}{1949}{1}.}
  \ref{Epstein}{Epstein,P. \ma{56}{1902}{615}.}
  \ref{Petersson}{Petersson, H. \ma {116}{1939}{401}.}
  \ref{Serre}{Serre,J-P. {\it A Course in Arithmetic}, Springer, New York,
  1973.}
  \ref{Schoenberg}{Schoenberg,B., {\it Elliptic Modular Functions},
  Springer, Berlin, 1974.}
  \ref{Apostol}{Apostol,T.M. \dmj {17}{1950}{147}.}
  \ref{Ogg2}{Ogg,A. {\it Lecture Notes in Math.} {\bf 320} (1973) 1.}
  \ref{Knopp}{Knopp,M.I. \dmj {45} {1978}{47}.}
  \ref{Knopp2}{Knopp,M.I. \invm {}{1994}{361}.}
  \ref{Knopp3}{Knopp,M.I. {\it Lecture Notes in Mathematics} {\bf 1383}(1989) 111 .}
  \ref{LandZ}{Lewis,J. and Zagier,D. \aom{153}{2001}{191}.}
  \ref{HandK}{Husseini, S.Y. and Knopp, M.I. {\it Illinois.J.Math.} {\bf 15} (1971)
  565.}
  \ref{Kober}{Kober,H. \mz{39}{1934-5}{609}.}
  \ref{HandL}{Hardy,G.H. and Littlewood, \am{41}{1917}{119}.}
  \ref{Watson}{Watson,G.N. \qjm{2}{1931}{300}.}
  \ref{SandC2}{Chowla,S. and Selberg,A. {\it Proc.Nat.Acad.} {\bf 35}
  (1949) 371.}
  \ref{Landau}{Landau, E. {\it Lehre von der Verteilung der Primzahlen},
  (Teubner, Leipzig, 1909).}
  \ref{Berndt4}{Berndt,B.C. \tams {146}{1969}{323}.}
  \ref{Berndt3}{Berndt,B.C. \tams {}{}{}.}
  \ref{Bochner}{Bochner,S. \aom{53}{1951}{332}.}
  \ref{Weil2}{Weil,A.\ma{168}{1967}{}.}
  \ref{CandN}{Chandrasekharan,K. and Narasimhan,R. \aom{74}{1961}{1}.}
  \ref{Rankin3}{Rankin,R.A. {\it Proc.Glas.Math.Assoc.} {\bf 1} (1953) 149.}
  \ref{Berndt6}{Berndt,B.C. {\it Trans.Edin.Math.Soc} {\bf 10} (1967) 309.}
  \ref{Elizalde}{Elizalde,E. {\it Ten Physical Applications of Spectral
  Zeta Function Theory}, \break (Springer, Berlin, 1995).}
  \ref{Allen}{Allen,B., Folacci,A. and Gibbons,G.W. \pl{189}{1987}{304}.}
  \ref{Krazer}{Krazer,A. {\it Lehrbuch der Thetafunktionen} (Teubner,
  Leipzig, 1903)}
  \ref{Elizalde3}{Elizalde,E. {\it J.Comp.and Appl. Math.} {\bf 118} (2000) 125.}
  \ref{Elizalde2}{Elizalde,E., Odintsov.S.D, Romeo, A. and Bytsenko, A.A and
  Zerbini,S.
  {\it Zeta function regularisation}, (World Scientific, Singapore, 1994).}
  \ref{Eisenstein}{Eisenstein,B. \jram {35}{1847}{153}.}
  \ref{Hecke}{Hecke,E. \ma{112}{1936}{664}.}
  \ref{Terras}{Terras,A. {\it Harmonic analysis on Symmetric Spaces} (Springer,
  New York, 1985).}
  \ref{BandG}{Bateman,P.T. and Grosswald,E. {\it Acta Arith.} {\bf 9} (1964) 365.}
  \ref{Deuring}{Deuring,M. \aom{38}{1937}{585}.}
  \ref{Guinand2}{Guinand, A.P. \qjm{6}{1955}{156}.}
  \ref{Guinand}{Guinand, A.P. \qjm {15}{1944}{11}.}
  \ref{Minak}{Minakshisundaram.}
  \ref{Mordell}{Mordell,J. \qjm {1}{1930}{77}.}
  \ref{GandZ}{Glasser,M.L. and Zucker,I.J. ``Lattice Sums" in Theoretical
  Chemistry, Advances and Perspectives, vol.5, ed. D.Henderson, New York,
  1986, 67  .}
  \ref{Landau2}{Landau,E. \jram{}{1903}{64}.}
  \ref{Kirsten1}{Kirsten,K. \jmp{35}{1994}{459}.}
  \ref{Sommer}{Sommer,J. {\it Vorlesungen \"uber Zahlentheorie} (1907,Teubner,Leipzig).
  French edition 1913 .}
  \ref{Reid}{Reid,L.W. {\it Theory of Algebraic Numbers}, (1910,MacMillan,New York).}
\end{putreferences}
\bye